\documentclass[11pt]{article}
\usepackage[font=small,labelfont=bf]{caption}
\usepackage{listings}
\usepackage{booktabs}
\usepackage{subcaption}
\usepackage{multirow}
\usepackage{colortbl}
\usepackage{amsmath,amssymb}
\usepackage[left=1in, right=1in, top=1in, bottom=1in]{geometry}
\usepackage{caption}
\usepackage{authblk} 

\usepackage{verbatim}
\usepackage{titling}
\usepackage{parskip}
\usepackage{todonotes} 

\usepackage{graphicx} 
\usepackage[english]{babel}
\usepackage[utf8]{inputenc}
\usepackage{slashed}
\usepackage{amsfonts}
\usepackage{amsthm}
\usepackage{amsmath,amssymb}
\usepackage{empheq}
\usepackage{dsfont}
\usepackage{xcolor}
\usepackage{hyperref}
\usepackage[normalem]{ulem}
\definecolor{L12_yellow}{RGB}{255, 232, 178}
\definecolor{L13_blue}{RGB}{178, 201, 255}
\definecolor{L1_green}{RGB}{156, 215, 160}
\definecolor{L3_red}{RGB}{255, 168, 168}

\usepackage{xcolor}

\usepackage{soul}

\begin{document}
\title{Inferring contact network characteristics from epidemic data via compact mean-field models}
\author{Andr\'es Guzm\'an$^1$, Federico Malizia$^1$, Gyeong Ho Park$^2$, Boseung Choi$^{2,3,4}$\thanks{cbskust@korea.ac.kr},\\ Diana Cole$^5$, Istv\'an Z. Kiss$^{1,6}$\thanks{istvan.kiss@nulondon.ac.uk}}

\affil[1]{Network Science Institute, Northeastern University London, London E1W 1LP, United Kingdom}
\affil[2]{Division of Big Data Science, Korea University Sejong Campus, Sejong, 30019, Korea}
\affil[3]{Biomedical Mathematics Group, Institute for Basic Science, Daejeon, 34126, Korea}
\affil[4]{College of Public Health, The Ohio State University, Columbus, OH 43210, USA}
\affil[5]{School of Engineering, Mathematics and Physics, University of Kent, Canterbury, United Kingdom}
\affil[6]{Department of Mathematics, Northeastern University, Boston, MA 02115, USA}

\date{\today}
\maketitle

\begin{abstract}
{Modelling epidemics using contact networks provides a significant improvement over classical compartmental models by explicitly incorporating the network of contacts. However, while network-based models describe disease spread on a given contact structure, their potential for inferring the underlying network from epidemic data remains largely unexplored. In this work, we consider the edge-based compartmental model (EBCM), a compact and analytically tractable framework, and we integrate it within dynamical survival analysis (DSA) to infer key network properties along with parameters of the epidemic itself. Despite correlations between structural and epidemic parameters, our framework demonstrates robustness in accurately inferring contact network properties from synthetic epidemic simulations. Additionally, we apply the framework to real-world outbreaks—the 2001 UK foot-and-mouth disease outbreak and the COVID-19 epidemic in Seoul— to estimate both disease parameters and network characteristics. Our results show that our framework achieves good fits to real-world epidemic data and reliable short-term forecasts. These findings highlight the potential of network-based inference approaches to uncover hidden contact structures, providing insights that can inform the design of targeted interventions and public health strategies.}
\\
\end{abstract}
\small{Keywords: Epidemics, Inference, Contact Networks}
\section{Introduction}

The spread of infectious diseases is inherently tied to the structure of human interactions. Network theory provides a powerful framework for understanding how diseases propagate by capturing the complex web of contacts between individuals \cite{kiss2017mathematics,pastor2015epidemic,newman_structure_2003,keeling_networks_2005,danon_networks_2011}. Studies have highlighted how structural properties such as heterogeneity \cite{moreno_epidemic_2002,chakrabarti_epidemic_2008,pastor2001epidemic,may2001infection,pastor2003epidemics}, communities \cite{Epi_community,Modularity1,nadini2018epidemic,griffin2012community}, clustering \cite{Epi_structured,smieszek2009models,miller2009percolation,coupechoux2014clustering}, and degree correlations \cite{bogua2003epidemic,wang2018edge,Epi_degree_correlation,chen2018predicting} play a significant role in shaping epidemic dynamics.
 
Epidemic models have traditionally been used to describe and predict disease spread based on assumptions about the underlying contact structure \cite{pastor2015epidemic,latora2017complex}. Ultimately, their applicability to real-world processes depends on the availability and quality of data \cite{malizia2022individual}. These models range in complexity, from classical mass-action approaches, where populations are assumed to mix homogeneously \cite{Kermack_original}, to sophisticated network-based frameworks that explicitly incorporate individual-level connectivity patterns \cite{volz2009epidemic,ball2008network,zhang2015modeling}. Simpler models offer easier tractability but may overlook key structural features, while more complex models provide richer descriptions but require more detailed input data \cite{mollison1995epidemic,gibson2018comparison,daunizeau2020reliability,malizia2022individual}. Finding a balance between these aspects is essential for effective epidemic modelling \cite{keeling2005networks}.

A particularly elegant and efficient modelling framework is the edge-based compartmental model (EBCM), which provides a compact yet powerful representation of epidemic processes on networks~\cite{miller2012edge,miller2013model,volz2011effects}. Unlike standard compartmental models where incorporating heterogeneity significantly increases model complexity, EBCM encodes network structure and characteristics through probability-generating functions, allowing epidemic dynamics to be described with only a few parameters and a reduced number of equations. 

In many real-world scenarios, direct measurements of contact networks are unavailable or incomplete. Although collecting data from contact networks is feasible in certain cases, such as sexually transmitted infections \cite{eng1997hidden,eames2003contact,rothenberg2003contact,garnett1993contact}, it remains challenging for respiratory diseases \cite{buckee2020improving,zubaydi2020using,kwok2019epidemic}. While epidemic models are often used to simulate outbreaks given a known network structure, inferring the structure of the contact network from observed epidemic data represents an equally important challenge \cite{vanli2024inference,viboud2018rapidd,cori2024inference,vasiliauskaite2022some,swallow2022challenges}. Since spreading dynamics inherently reflect network properties, they can be used to extract valuable information about the underlying structural information. Various methods have been proposed to reconstruct networks from data, including likelihood-based optimization approaches \cite{InferTopology_2011_Shandilya, nitzan_2017, netrapalli_learning_2012,malizia2024reconstructing, rodriguez_uncovering_2014} and Bayesian inference techniques \cite{groendyke_bayesian_2011, britton_bayesian_2002, lewis_episodic_2008, demiris_bayesian_2005, prasse_exact_2019}. However, these methods often require detailed temporal data or strong prior assumptions, making them difficult to apply in real-world epidemic surveillance \cite{gallo2022lack}.

An alternative approach, Dynamical Survival Analysis (DSA), has been introduced to estimate epidemic parameters using infection and recovery time distributions \cite{SDS}. Originally developed for mass-action models \cite{di_lauro_dynamic_2022, rempala_dynamical_2023}, DSA was recently extended to network-based models \cite{kiss_towards_2023}, enabling parameter estimation while incorporating some aspects of network structure. However, existing applications remain limited in their ability to fully capture the heterogeneity of contact networks.

In this paper, we integrate the DSA approach with the edge-based compartmental model (EBCM) \cite{miller2012edge} to develop a Bayesian framework for inferring both disease and network parameters from epidemic data. This extends previous works \cite{kiss_parameter_2023}, shifting from identifiability analysis to active inference in both synthetic and real-world scenarios. The manuscript is structured as follows: Section \ref{sec:EBCM_model} introduces the EBCM framework, and Section \ref{sec:inf_methods} details the Bayesian inference procedure. Section \ref{sec:Results} presents validation on synthetic and real data, specifically the first wave of COVID-19 in Seoul and the 2001 foot-and-mouth disease epidemic in the UK. Finally, Section \ref{sec:Discussion} discusses the implications of our findings.

\section{Methods}
\label{sec:methods}

In this section, we outline the methodologies that form the foundation of our inference framework. First, we introduce the edge-based compartmental model (EBCM), which provides a compact representation of SIR processes on networks. This model serves as the backbone for describing the epidemic dynamics in structured populations.

Next, we present the complete inference process, detailing how these methods are integrated to estimate both epidemic and network parameters from observed outbreak data. Specifically, we employ Dynamic Survival Analysis (DSA) to construct the likelihood function, leveraging its ability to handle censored and aggregated epidemic data. Moreover, we describe the Robust Adaptive Metropolis (RAM) algorithm, a Markov Chain Monte Carlo (MCMC) technique designed for efficient exploration of the parameter space. RAM adapts to the local structure of the posterior distribution, improving convergence and robustness in high-dimensional settings. Together, these methods form a comprehensive framework for inferring epidemic dynamics and network structures from real-world outbreak data.

\subsection{Edge-based compartmental model}
\label{sec:EBCM_model}

We consider a Susceptible-Infected-Recovered (SIR) epidemic process, where individuals can be in one of three states: susceptible ($S$), infected ($I$), or recovered ($R$). Infection occurs at rate $\beta$ along a link between a susceptible and an infected node, while infected nodes recover independently of the network at rate $\gamma$. In this study, we employ the Edge-Based Compartmental Model (EBCM) \cite{miller2012edge}, which provides a compact and analytically tractable representation of epidemic dynamics on contact networks. The EBCM assumes that disease transmission occurs on a network generated by the configuration model (CM) \cite{molloy1995critical,newman2001random}, which is characterized by a degree distribution $P(k)$. The key idea behind EBCM is to track the probability that a randomly chosen node remains susceptible rather than explicitly tracking individual infection events.  

A central variable in the model is $\theta(t)$, defined as the probability that a randomly selected neighbor of a test node $u$ has not transmitted the disease to $u$ by time $t$. From now on, we omit the obvious time dependence. Given that a node $u$ has degree $k$, the probability that it remains susceptible is $s_u(k,\theta) = \theta^k$. Thus, the overall fraction of susceptible nodes in the population is given by:

\begin{equation}
S(t) = \sum_k P(k) \theta^k = \Psi(\theta),
\end{equation}
where $\Psi(\theta)$ represents the probability generating function (PGF).  

If a fraction $\rho$ of the population is initially infected at $t=0$, we modify this expression as $
S(t) = \hat{\Psi}(\theta) = \sum_k P(k) S(k,0) \theta^k$,
where $S(k,0) = 1 - \rho$.  
To fully characterize the system, we decompose $\theta$ into three probabilities, namely $\theta = \psi_S + \psi_I + \psi_R$,
where $\psi_S$, $\psi_I$, and $\psi_R$ represent the probabilities that a randomly chosen neighbor of node $u$ is in the susceptible, infected, or recovered state, respectively, and has not yet transmitted the infection to $u$. With these ingredients, we can write $\dot{\theta} = \beta \psi_I$, and we can express  $\psi_R=\frac{\gamma}{\beta}(1-\theta)$, and  $\psi_S = \hat{\Psi}'(\theta)/\langle k \rangle$ where $\hat{\Psi}'(\theta)$ is the derivative of the PGF with respect to $\theta$, while $\langle k \rangle = \sum_k k P(k)S(k,0)$, corresponding to the derivative of the PGF evaluated at $\theta = 1$. 
By incorporating these expressions, the evolution of the system is governed by:
\begin{equation}
    \begin{split}
    \frac{d\theta}{dt} &= -\beta \theta + \beta \psi_{S}(0) \frac{\hat{\Psi}'(\theta)}{\langle k \rangle} + \gamma (1 - \theta) + \beta \psi_{R}(0), \\
    \frac{dR}{dt} &= \gamma(1 - S - R), \quad S = \hat{\Psi}(\theta).
    \end{split}
    \label{eq:EBCM}
\end{equation}

Typically, we assume $\psi_R(0) = 0$ and $\psi_S(0) = 1 - \rho$. Solving Eqs.~\eqref{eq:EBCM} provides the evolution of $S(t)$, $I(t)$, and $R(t)$. Moreover, the basic reproductive number ($R_0$) of the EBCM is defined as 
\begin{equation}
\label{eq:R_0}
    \begin{array}{l}
   R_0 = \dfrac{\beta}{\beta+\gamma}\dfrac{\langle k^2 \rangle - \langle k \rangle}{\langle k \rangle},
    \end{array}
\end{equation}
where $\langle k^2 \rangle - \langle k\rangle = \sum_{k}k(k-1)P(k)S(k,0)$, corresponding to the derivative of the PGF evaluated at $\theta = 1$. For simplicity, from now on, we apply a change of variable $\mu \equiv \langle k \rangle$. 

\begin{table}[t!]
    \centering
    \begin{tabular}{ccc}
        \toprule
                        & \text {Poisson}
                        & \text {Negative Binomial}
                        \\[5pt]
        \midrule
        \text {Parameter(s)}    
                        
                        & $\mu$   
                        & $(r, \mu)$ 
                        \\[3pt]
        $\Psi(x)$               
                        & $e^{\mu(x - 1)}$  
                        & $\bigg(\frac{r}{r+\mu(1-\theta)}\bigg)^{r}$ 
                        \\[3pt]
        $\Psi'(x)$               
                        & $\mu e^{\mu(x - 1)}$ 
                        & $\mu \bigg(\frac{r}{r+\mu(1-\theta)}\bigg)^{r+1}$ 
                        \\[3pt]
                        \\
        \bottomrule
    \end{tabular}
    \caption{Details of the probability generating functions used throughout the paper. The parameter $\mu$ for both distributions corresponds to the average degree given by $\sum_{k}kP(k)$}.
        \label{tab:PGF}
\end{table}

In this study, we consider two different degree distributions, which are summarized in Table~\ref{tab:PGF} along with their parameters and probability generating functions. The Poisson distribution is characterized by a single parameter $\mu$, which defines both its mean and variance, resulting in a relatively homogeneous degree distribution. In contrast, the Negative Binomial distribution, parametrized by $r$ (number of successes) and $\mu$ (mean number of failures before $r$ successes), provides greater flexibility, allowing for the modelling of both homogeneous and heterogeneous network structures.

                        


\subsection{Statistical Inference framework}
\label{sec:inf_methods}

Accurate parameter estimation in epidemic modelling usually relies on optimizing a likelihood function that reflects both the underlying transmission dynamics and the nature of the available data. A common approach involves fitting model-generated epidemic curves to observed data by minimizing discrepancies between them. However, this method is highly sensitive to noise, biases, and incomplete datasets, which can compromise inference accuracy. To address these challenges, we employ the \textit{Dynamic Survival Analysis} (DSA) framework~\cite{KhudaBukhsh2020DSA,DiLauro2022NonMarkovDSA,Vossler2022Ebola,khuda_bukhsh_2022_projecting}, which provides a more robust approach by directly incorporating individual transition times between epidemic states into the likelihood function.

DSA was developed to overcome the limitations of traditional inference methods in infectious disease epidemiology by integrating dynamical systems theory with survival analysis techniques. Unlike conventional approaches that rely on aggregate epidemic curves, DSA leverages the mean-field ordinary differential equations (ODEs) governing population-level dynamics to model the probability distributions of transition times, such as the time of infection or recovery. This formulation allows DSA to construct likelihood functions for individual-level trajectories, making it particularly effective in handling censored, truncated, or incomplete data. In this framework, the susceptible fraction of the population, $S(t)$, is reinterpreted as a survival function, satisfying $S(0) = 1$. More generally, when a fraction $\rho$ of individuals is initially infected, we introduce three rescaled survival functions, which are defined as 

\begin{equation}
\label{eq:rescaled_survival_functions}
    \tilde{S}(t) = \frac{S(t)}{1-\rho}=\Psi(\theta),  \hspace{0.5cm}  \tilde{I}(t) = \frac{I(t)}{1-\rho},  \hspace{0.5cm}   \mbox{~and~}   \hspace{0.5cm}  \tilde{R}(t) = \frac{R(t)}{1-\rho}.
\end{equation}
By substituting Eq.~\eqref{eq:rescaled_survival_functions} in the system of equations for the EBCM, as given by Eqs.~\eqref{eq:EBCM}, we have
\begin{equation}
\begin{split}
        &\Dot{\tilde{S}}(t) = \frac{d\tilde{S}}{d\theta}\frac{d\theta}{dt} 
     = \Psi'(\theta)\Dot{\theta}
     = \Psi'(\theta) \Bigg[-\beta \theta + \beta (1-\rho)\frac{\hat{\Psi}'(\theta)}{\langle k \rangle}+\gamma(1-\theta)\bigg]. \\
    &\dot{\tilde{R}}(t) = \gamma \tilde{I}(t) \hspace{0.5cm} \mbox{~and~} \hspace{0.5cm} \tilde{I}(t) =1/(1-\rho)-\tilde{S}(t)-\tilde{R}(t),
\end{split}
\label{eq:EBCM_rescaled}
\end{equation}
where, at $t=0$ we have $\tilde{S}(0)=1$, $\tilde{I}(0)=\rho/(1-\rho)$ and $\tilde{R}(0)=0$.

DSA interprets the susceptible curve as an improper survival function representing the time of infection for a randomly chosen initially susceptible individual. That is, \( \tilde{S}(t) = \mathsf{P}(T_I > t) \), where the random variable \( T_I \) denotes the infection time. The density function of \( T_I \) is given by \( -\Dot{\tilde{S}}(t) \), which is improper since \( \lim_{t\to \infty} \tilde{S}(t) = \mathsf{P}(T_I = \infty) > 0 \). We define \( \mathsf{P}(T_I = \infty) = 1 - \tau \), where \( \tau \) represents the final epidemic size. To obtain a proper survival function, we condition it on a final observation time \( T \in (0, \infty) \) and the final epidemic size \( \tau \) at time \( T \). The resulting probability density function \( f_{\tau}(t) \) on the interval \( [0, T] \) is then given by:

\begin{align}
    f_\tau (t) = - \frac{\dot{\tilde{S}}(t)}{\tau}. 
    \label{eq:density_ti}
\end{align}

Note that DSA does not require knowledge of recovery times. However, if these times are available, they can be incorporated to enhance the quality of inference. Let \( T_R \) represent the time of recovery of an infected individual. Given the infection time \( T_I \), the infectious period \( T_R - T_I \) follows an exponential distribution with rate \( \gamma \). Using equation~\eqref{eq:density_ti} and the density of the infectious period, we can define the density of the recovery time \( T_R \) as:

\begin{equation}
    g(t) = \int_0^{t} f_{\tau}(u) \gamma e^{-\gamma (t-u)}\mathrm{d}u.
    \label{eq:density_tr1}
\end{equation}

Equation~\eqref{eq:density_tr1} represents the convolution of the density of the infection time \( f_\tau (t) \) and an exponential distribution with rate \( \gamma \), corresponding to the density of the infectious period. In practice, solving the system of ODEs~\eqref{eq:EBCM_rescaled} with respect to the observed recovery times is computationally more convenient.

Finally, the normalized density of the recovery time is given by:

\begin{equation}
    \tilde{g}(t) = \frac{g(t)}{\int_0 ^{T}g(t)\mathrm{d}t}.
    \label{eq:density_tr}
\end{equation}

One of the advantages of the DSA method is the ability to build various likelihood functions based on the observed data. Let $N$ be the size of the population and $M$ be the initial number of infected individuals at the beginning time $t=0$, and usually $N>>M$. We have $K$ individuals out of $N$, who are infected by time $T$ and $t_1, t_2..., t_K$  represent the time of infection for each infected individual. If $L$ individuals have recovered by time $T$ out of a total of $K + M$ infected, let $r_1, r_2,..., r_L$  denote the time of recovery. If we can observe the time of infection and recovery exactly for each individual, then we can define the infectious period, $w_i = r_i - t_i$ or $w_i = r_i$ for initially infected, $i =1,2,\dots, L$ respectively. Also, we could have $\tilde{L}$ infected individuals who have not recovered by time $T$, $\epsilon_j = T-t_j$ or $\epsilon_j = T$, $j=1,2,\ldots, \tilde{L}$, denote the censored infectious period respectively.

Given random samples of time of infection, $t_1, t_2,..., t_K$,the log likelihood function is given by:
\begin{align}
	\ell_{1} = K\log(\tau) + (N-(M+K))\log(1-\tau) + \sum_{i=1}^{K} f_\tau (t_i).
    \label{eq:l1}
\end{align}
The log likelihood function for the infectious period, $w_1, w_2,..., w_L$ is given by 
\begin{align}
	\ell_{2} = L\log(\gamma) - \gamma \sum_{j=1}^{L} w_j.
    \label{eq:l2}
\end{align}
The log likelihood function for the time of recovery, $r_1, r_2,..., r_L$ is given by 
\begin{align}
	\ell_{3} = \sum_{j=1}^{L} \tilde{g}(r_j).
    \label{eq:l3}
\end{align}
Finally, we can define the log likelihood for the censored infectious period, $\epsilon_1, \epsilon_2,\ldots, \epsilon_{\tilde{L}}$;
\begin{align}
	\ell_{4} = -\gamma \sum_{k=1}^{\tilde{L}} \epsilon_k.
    \label{eq:l4}
\end{align}

In some cases, we only know the number of infected individuals, $K$, by a given time, $T$, but lack information about the total population size $N$. This situation is common in real epidemic scenarios, where data about the segment of the population at risk of infection is often unavailable. For such cases, the likelihood $\ell_{1}$ can be reformulated as: $\ell_{1} = \sum_{i=1}^{K} f_\tau (t_i)$, excluding terms related to the total population size. With this formulation, we can analyze the dynamics of the proportion of infected, susceptible, and recovered individuals. Furthermore, it is possible to estimate an effective population size using the following equation

\begin{equation}
    N_{eff} = \frac{K}{1-S_T}.
\end{equation}

Depending on the data and particular context, we can use any of the four likelihoods described above or a combination of them. For example, suppose we just observe $K$ infections before a cut-off time $T$, then we could just use $\ell_{1}$  ~\eqref{eq:l1}. If, besides the infection times, we also observe $L$ recovery times but not the specific nodes who underwent these changes, then we could use the combination of $\ell_1$ + $\ell_3$, see Eqs.~ ~\eqref{eq:l1} and ~\eqref{eq:l3}. However, if the data is available at the node level and we have pairs of infection and recovery times for each node, we could use the combination $\ell_1$+ $\ell_2$, see Eqs. ~\eqref{eq:l1}~\eqref{eq:l2}, since we can calculate the infectious period for each node. 
The likelihood functions described above do not explicitly incorporate the network parameters, making it impossible to derive closed-form solutions for their maximum likelihood estimates. To address this limitation, we adopt a Bayesian approach, which allows us to sample from the posterior distribution and derive point estimates from it. To construct the sample, we employ a Markov Chain Monte Carlo (MCMC) method, specifically the Robust Adaptive Metropolis (RAM) algorithm. The RAM algorithm is more efficient than the standard Metropolis-Hastings algorithm \cite{RAM}, as it dynamically adjusts the variance-covariance matrix of the proposal distribution to maintain an optimal acceptance rate during the Metropolis steps. We assign vaguely informative prior distributions (little information is given about the parameters to be estimated) to the model parameters: a Gamma distribution, $\mbox{Gamma}(a, b)$, for the $\beta$ and $\gamma$ parameters, which is defined as 
\begin{equation}
    \begin{array}{l}
    \mbox{Gamma}(x,a,b) \sim \dfrac{b^a x^{a-1} e^{-bx}}{\Gamma(a)},
    \end{array}
\end{equation}

where $\Gamma(a)$ represents the gamma function.
Additionally, for the parameter $\rho$ in the SIR model, we assume a Beta distribution, $\mbox{Beta}(a, b)$, where, in this context, $a=1$ and $b=1$ are chosen from a uniform distribution.
Additionally, we assign a non-informative $\mbox{Gamma}$ prior to the parameters of the degree distribution based on the support of these parameters.

\begin{figure}[t!]
	\centering
	\includegraphics[width=0.9\textwidth]{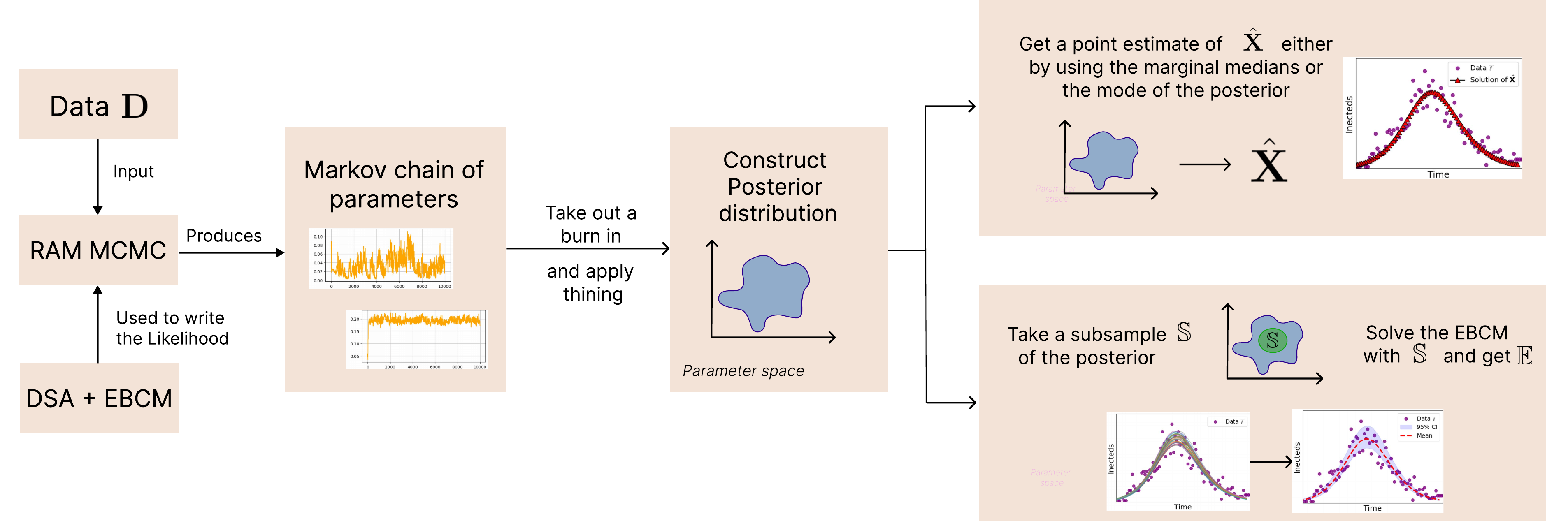}
\caption{\textbf{Graphical representation of the inference process.} The procedure starts with epidemic data $\mathbf{D}$ as input for the Robust Adaptive Metropolis (RAM) algorithm, which employs the likelihood function derived from the DSA framework (see main text). The RAM algorithm generates a Markov chain of parameter samples, forming the posterior distribution. From this distribution, we obtain parameter estimates using two approaches: credible intervals and point estimates, as detailed in the main text.}
	\label{fig:diagram}
\end{figure}

Given a dataset $\mathbf{D}$, our goal is to fit the data with the EBCM, which is characterized by the set of parameters $\mathbf{X} = (\beta, \gamma, \rho, \mathbf{\Delta})$. Here, we use the vector $\mathbf{\Delta}$ to represent the parameters of the probability generating function of the network's degree distribution. Our objective is to infer the parameter set $\mathbf{X}$ such that the output from the EBCM matches the observed data. To achieve this, we utilize the RAM algorithm to generate one or more chains for the values of $\mathbf{X}$, which allows us to construct a sample of the posterior distribution. From this sample, we remove the burn-in and apply a thinning procedure to reduce autocorrelation. The inference scheme is illustrated in the left panels of Fig.~\ref{fig:diagram}. 

Once a posterior distribution is obtained, we can find a point estimate for each parameter. This can be done by either calculating the \textit{marginal mean}, \textit{marginal median} or taking the \textit{joint mode} of the posterior distribution, i.e., the point of highest probability in the full posterior. As shown in the top right panel in Fig.~\ref{fig:diagram}. Either of these can be used in conjunction with the EBCM to produce one single epidemic curve that can be compared to the data. However, to generate a credible interval around this single epidemic curve, we take a subsample $\mathbb{S}$ from the full posterior and use each element to solve the EBCM, thereby generating a set of solutions, as shown in the bottom right panel of Fig.~\ref{fig:diagram}. These can used to generate the desired credible interval over a specified period of time. 

In subsequent sections, we show the application of this workflow to estimate network and epidemic dynamics parameters for two different scenarios. First, by the use of synthetic data corresponding to a controlled case where ground truth is available. Second, we consider two different datasets corresponding to real-world epidemics. 

\label{sec:Results}
\section{Inference from synthetic data}
\label{sec:Results}

 We begin by analysing synthetic data to verify the method’s accuracy and effectiveness in a controlled environment. First, we consider 100  realizations of Gillespie simulations~\cite{Gillespie1976} with $\beta=0.2$, $\gamma=1$, on networks with $10^4$ nodes exhibiting Poisson ($POI(\mu=10)$) and Negative Binomial ($NB(\mu=10,r=1)$) degree distributions. Since in a real epidemic process, complete data are rarely available, we consider four possible scenarios which reflect varying data availability, which are:
\begin{itemize}
    \item $\ell_1+\ell_2$:  infection times $(t_1,t_2,\ldots,t_K)$ and infection periods  $(w_1,w_2,\ldots,w_L)$ 
    \item $\ell_1+\ell_3$: list of infection times $(t_1,t_2,\ldots,t_K)$ and a decoupled list of recovery times $(r_1,r_2,\ldots,r_L)$.
    \item $\ell_1$:  infection times only $(t_1,t_2,\ldots,t_K)$.
    \item $\ell_3$:  recovery times only $(r_1,r_2...r_L)$.
\end{itemize}

Following the procedure outlined in Section \ref{sec:inf_methods}, we fit our models to all stochastic realizations. Initially, we consider the case where the data are fitted using the correct model. Specifically, data from simulations on networks with a Poisson degree distribution were fitted using the Poisson (Poi) model in the EBCM; we refer to this scenario as Poisson-Poi. Similarly, a match with the Negative Binomial is denoted by Negbin-NB. We also consider model mismatch, where data from networks with a Poisson degree distribution were fitted using the Negative Binomial (NB) model (Poisson-NB), and vice-versa, that is Negbin-Poi.

\begin{figure}[t!]
	\centering
	\includegraphics[width=1\textwidth]{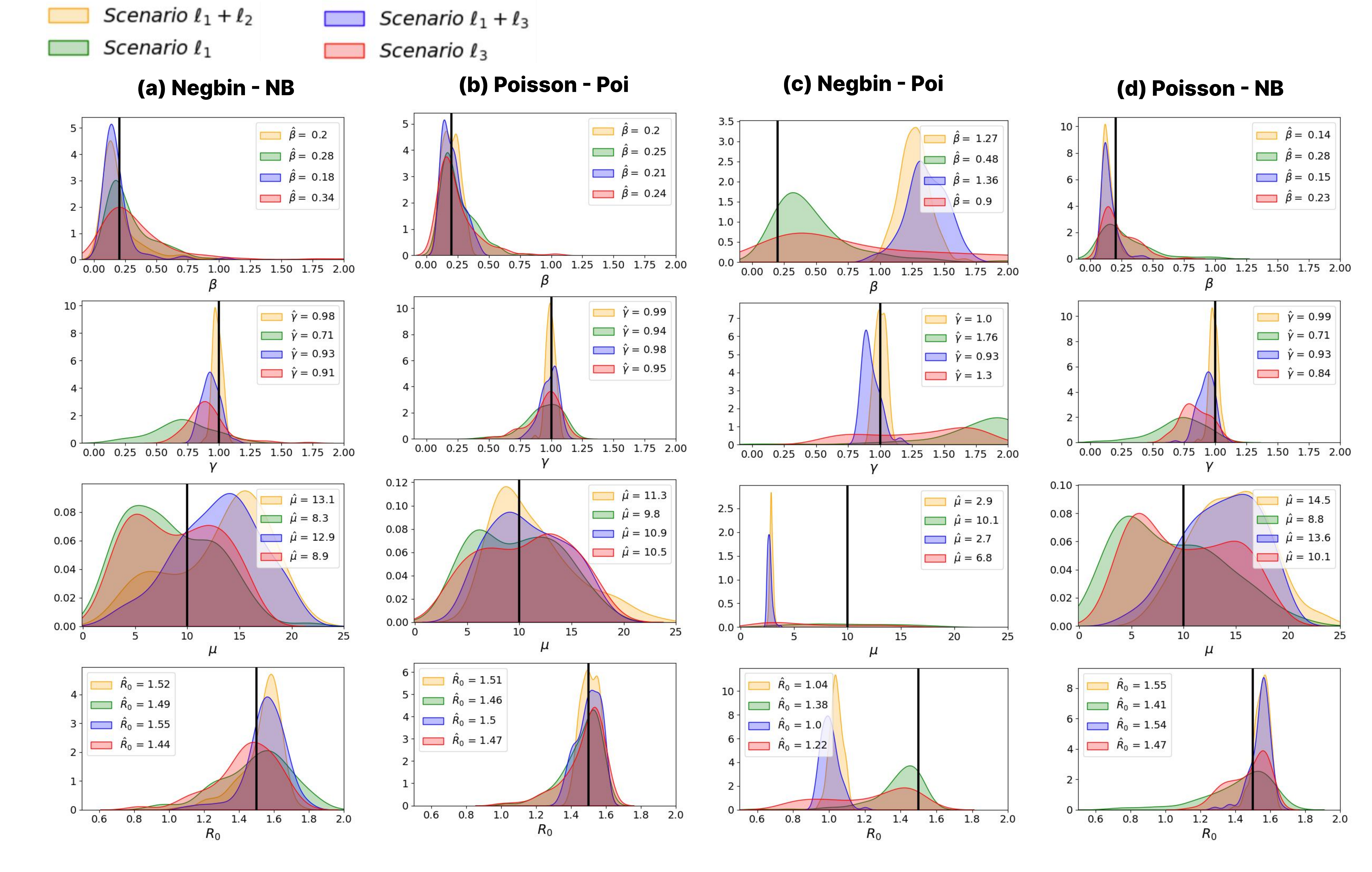}
	\caption{\textbf{Distributions of the marginal medians for model parameters and the basic reproduction number inferred from the DSA framework.} Each row represents the distribution of a different inferred parameter. The synthetic data were generated using either a Poisson or a Negative Binomial degree distribution, and inference was performed using both distributions for comparison. Each column corresponds to different inference cases: 
(a) and (b) show results where the data were fitted with the same degree distribution used for generation—(a) for Negative Binomial and (b) for Poisson.   
(c) and (d) show cases where the data were fitted using the opposite degree distribution—(c) for Negative Binomial data fitted with a Poisson model and (d) for Poisson data fitted with a Negative Binomial model.  
This analysis highlights the impact of assuming different degree distributions on parameter inference. The results are obtained by fitting 200 distinct realizations of Gillespie simulations with parameters $\beta = 0.2$, $\gamma = 1$, and an initial proportion of infected individuals set to $10^{-4}$. Networks with a Negative Binomial (Negbin) degree distribution were generated using parameters $\mu = 10$, $r = 1$, while networks with a Poisson degree distribution were generated with $\mu = 10$ }   
	\label{fig:Sim_study}
\end{figure}

For all four possible combinations (Negbin-NB, Poisson-Poi, Negbin-Poi, Poisson-NB), we used the four different scenarios discussed above. Then, for each stochastic realization, we construct a sample of the posterior distribution and calculated the marginal median of $\beta$, $\gamma$, and $\mu$, and computed the basic reproduction number as defined in Eq.\eqref{eq:R_0}. The point estimates for $\hat{\beta}$, $\hat{\gamma}$, $\hat{\mu}$, $\hat{R}_0$ are obtained as the mean values of the distributions of the marginal medians. In Fig.~\ref{fig:Sim_study}, we show the density of the marginal medians and their mean, i.e., point estimates, for these four parameters and for both model match and mismatch. Additionally, we present these results for the four different data availability scenarios presented above.

As expected, the correct matches, Negbin-NB, first column,  and Poisson-Poi, second column, yield accurate estimations for all parameters and for all different likelihoods. In the presence of data and model mismatch, we observe that fitting Poisson data with the Negative Binomial model (fourth column) often yields reasonable results. This can be attributed to the flexibility of the Negative Binomial distribution, i.e. one more free parameter compared to Poisson, which allows it to capture both homogeneous and heterogeneous degree distributions. In contrast, fitting data generated with the Negative Binomial with the Poisson model underperforms. This is especially evident in the estimates for $\beta$ and $\mu$ in the first and third rows, where the EBCM with a Poisson degree distribution tends to overestimate $\beta$ and underestimate $\mu$.

It is noteworthy that the mismatched Negbin–Poi model produced better estimates for $\beta$ and $\mu$ in the $\ell_1 + \ell_3$ and $\ell_3$ scenarios. However, even in these cases, $\gamma$ was substantially overestimated. These qualitative observations are further supported by the larger bias and mean squared error (MSE) of the estimates, as detailed in Supplementary Material (SM).  For the estimation of $\gamma$, scenario $\ell_1 + \ell_2$ uses the infectious periods as data. Since the infectious period directly reflects $\gamma$ (the inverse of the mean infectious period), this scenario yields highly accurate estimates, even for mismatched models and data. By contrast, estimates of $\gamma$ exhibit a larger bias in mismatched scenarios $\ell_1 + \ell_3$ and $\ell_3$, where the recovery times are used instead, but they do not align with the infection times.
\begin{figure}[t!]
	\centering
	\includegraphics[width=0.65\textwidth]{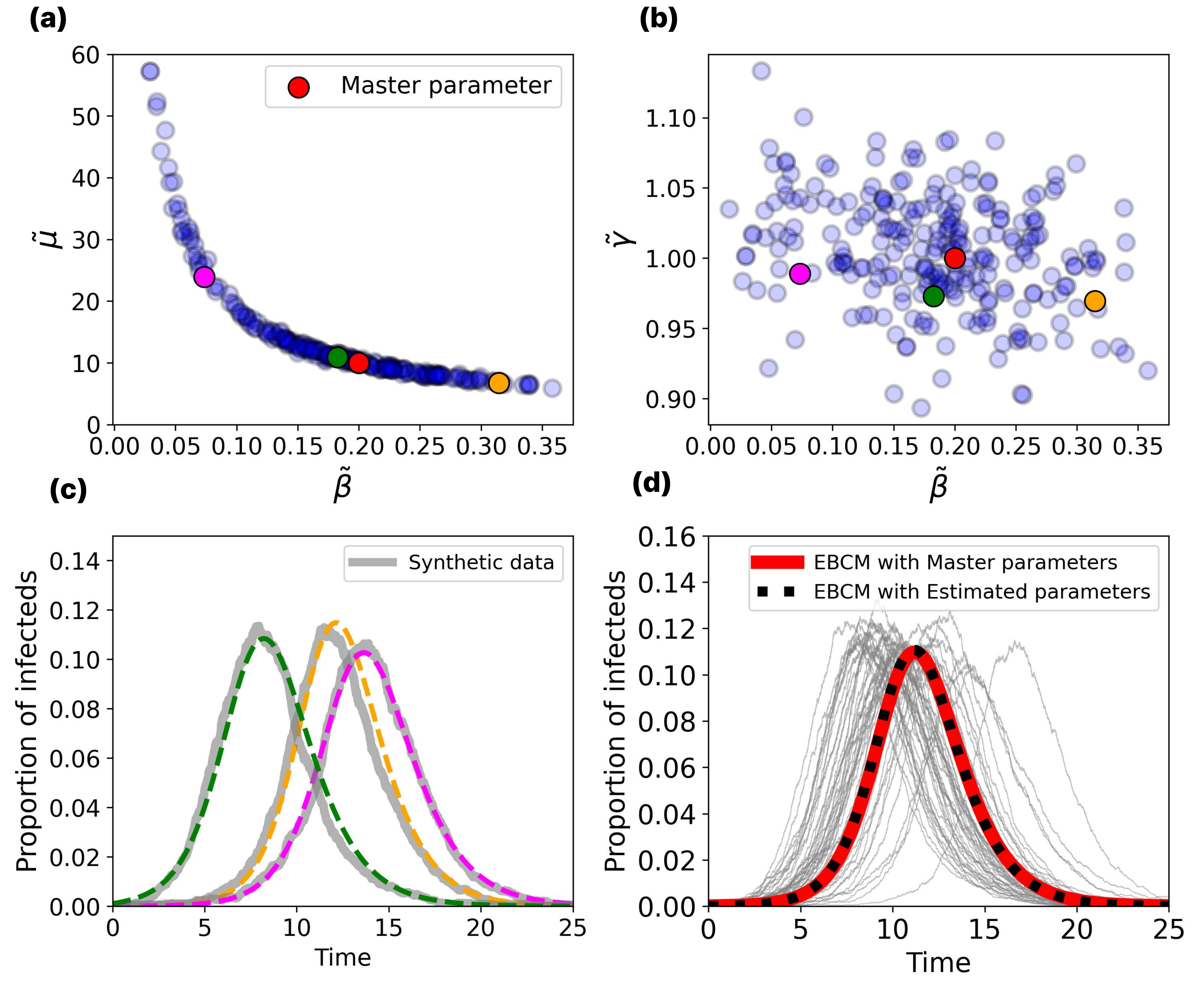}
	\caption{\textbf{DSA parameters estimation for each synthetic dataset.} (a) Projection of the 250 median estimates from the marginal posterior distributions of each dataset in the $\beta$–$\mu$ space. (b) Projection of the median estimates in the $\gamma$–$\beta$ space. (c) Epidemic curves from three selected datasets (gray lines) alongside the corresponding curves obtained by solving the EBCM using the median parameter estimates (pink, green, and orange points in panels (a) and (b)). Despite variations in parameter estimates, the model accurately reproduces the observed epidemic dynamics of the data. Panel (d) shows 50 of the 250 datasets (gray lines) with the solution of the EBCM obtained using the master parameters (dotted black line) and the solution from the final point estimates from the marginal posteriors (red line). Results are based on 250 synthetic epidemic datasets generated on a network with a Poisson degree distribution. The true parameters used were $\beta = 0.2$, $\gamma = 1$, $\mu = 10$, and $\rho = 1 \times 10^{-4}$. For inference, the cut-off time was set to $T = 25$.}
	\label{fig:Results_1}
\end{figure}
Interestingly, in the mismatch cases, in particular, see the panel on the third row and third column in Fig.~\ref{fig:Sim_study}, we note that the estimates of $\mu$ based on scenario $\ell_1 + \ell_2$ perform poorly when compared to estimates based on scenarios $\ell_1$ and $\ell_3$. This, at first, is surprising since $\ell_1 + \ell_2$ provides the most complete data. However, due to the model mismatch manifesting itself mainly in the infection part, this is likely to lead to a better estimation of $\gamma$, which in turn has a negative effect on the estimation of  $\beta$ and $\mu$. In contrast, in scenarios $\ell_1$ and $\ell_3$, the estimation of $\mu$ improves at the expense of obtaining less accurate estimates of $\gamma$. These observations are further supported by the MSE values presented in SM, where this analysis is discussed in detail. 

Even though accurate and precise estimations can be obtained using the marginal medians of the posterior distributions, it is important to recognize that the posterior distribution is multivariate. This means that the marginals are not always fully representative of the entire distribution, especially when there is a correlation between parameters. When used for inference,  network-based mean-field models have been noted to exhibit correlations between the infectivity $\beta$ and parameters of the degree distribution, such as the average degree $\mu$ \cite{kiss_towards_2023}. For this reason, we now investigate the projections of the posterior distribution. 
We do this for Poisson-Poi case, and we consider the scenario where the data is complete, i.e., $\ell_1 + \ell_2$. In Fig.~\ref{fig:Results_1}, 
 we show the projection of the medians of the marginals in the $(\beta,\mu)$ space.

In Fig.~\ref{fig:Results_1}(a), we observe a strong inverse correlation between the infection rate $\beta$ and the average degree $\mu$. 
Although we observe a considerable variance in both the average degree and the infection rate, it's important to highlight that most of the estimates of parameters are around the master parameter used in the simulation. Furthermore, no correlation is observed when exploring the $(\beta,\gamma)$ parameter space. As expected, the correlations between parameters can be mitigated by fixing either the infection rate or the average degree. This approach leads to a posterior sample without long tails, resulting in more accurate estimates. Further discussions on the process of fixing parameters are provided in SM. 


Up to this point, we have focused on parameter estimation with their validity based on comparison to their true values. However, another important consideration is to assess how the epidemic curve, such as new infecteds or prevalence, generated using the EBCM with the point estimates compares to the original outbreak data. To address this, we consider three different stochastic realizations of the epidemic, see grey curves in Fig.~\ref{fig:Results_1}(c). We fit each of these separately with the corresponding point estimates reported as the pink, green, and orange points in  Fig.~\ref{fig:Results_1}(a) and  Fig.~\ref{fig:Results_1}(b). The EBCM with these point estimates leads to excellent agreement with the original stochastic realizations, see Fig.~\ref{fig:Results_1}(c).

\section{Inference from real-world data}
\label{subsec:Real_data}

In this section, we further validate our methodology by analyzing real epidemic data from two outbreaks. Based on the general procedure presented in Section \ref{sec:inf_methods}, we apply it to two real epidemic datasets. The first one is the 2001 Foot-and-Mouth Disease (FMD) epidemic in the United Kingdom, which involved a highly contagious virus affecting farm animals. The second dataset captures the first wave of COVID-19 in Seoul, South Korea, documenting the onset of symptoms and confirmation of infection for approximately 500 individuals over the first 82 days following the appearance of the initial confirmed case.

For both datasets, we fit the EBCM using a probability-generating function corresponding to a Negative Binomial degree distribution, chosen for its flexibility in modelling both homogeneous and heterogeneous contact patterns. We adopt non-informative prior distributions, assigning $\beta$, $\gamma$, $\mu$, and $r$ a $\mbox{GAMMA}(a,b)$ prior, where $a$ is randomly selected from the parameter space and $b$ is fixed at $10^{-4}$. The initial number of infected individuals follows a $\mbox{BETA}(1,1)$ prior, representing a uniform distribution and reflecting complete uncertainty about the starting conditions. A key distinction from our analysis of synthetic data is that the total population size is unknown. As discussed in Section~\ref{sec:inf_methods}, one approach is to estimate an effective population size using $N_{eff} = K / (1 - S[T])$ \cite{DiLauro2022NonMarkovDSA}. To obtain a denser posterior distribution, we performed multiple runs of the RAM algorithm, varying the initial conditions of the chain for each run. Consequently, we find point estimates for each parameter and evaluate the accuracy of the EBCM predictions by comparing them to real-world epidemic data. To quantify the discrepancy, we use the mean squared error (MSE), defined as  
\begin{equation}
    MSE = \frac{1}{T} \sum_{d=0}^{T} (\hat{J}(d) - J(d))^2,
\end{equation}  
where $d$ denotes the day index, $T$ is the cut-off time, $\hat{J}(d)$ represents the incidence predicted by the EBCM, and $J(d)$ corresponds to the epidemic incidence observed in the real-data.  

Here, we compare the predictions of the EBCM with those obtained using the standard mass-action (MA) SIR model \cite{SDS}. The MA model is incorporated into the inference framework described in Section \ref{sec:inf_methods} and serves as a benchmark for epidemic curve predictions. The governing equations for the SIR MA model are given by:  

\begin{equation}
    \dot{s}_t = -\sigma s_t\iota_t,  \hspace{1cm} 
    \dot{\iota}_t = \sigma s_t\iota_t  - \gamma\iota_t, \hspace{1cm}
    \dot{r}_t = \gamma\iota_t,
\end{equation}  

where $\gamma$ is the recovery rate, and $\sigma$ is the infection rate. It is important to note that $\sigma$ differs from the infection rate $\beta$ in the EBCM: while $\beta$ represents the per-contact transmission rate, $\sigma$ describes the infection rate per infectious individual.  

Unlike the EBCM, which explicitly accounts for the network structure, the MA model assumes homogeneous random mixing, meaning it does not incorporate any connectivity patterns. As a result, while the EBCM enables inference of both the epidemic dynamics and the underlying contact structure, the MA model can only be used to predict the epidemic curve. Consequently, our comparison is limited to the epidemic trajectory rather than network-related properties. For parameter inference, we follow the same numerical procedure as outlined earlier.

\subsection{Foot-and-Mouth disease data}
\label{subsubsec:FMD}

In this section, we analyze the Foot-and-Mouth Disease (FMD) dataset, which provides daily incidence data, $J(d)$, representing the number of newly infected individuals per day over a 200-day period. We focus on the first 82 days, corresponding to the initial wave of the outbreak. In the DSA framework, performing inference requires information about the temporal distribution of infections. To address this, we assume that infection times within each day are uniformly distributed and, based on this assumption, consider only the $\ell_1$ likelihood.  

\begin{figure}[t!]
	\centering
	\includegraphics[width=0.75\textwidth]{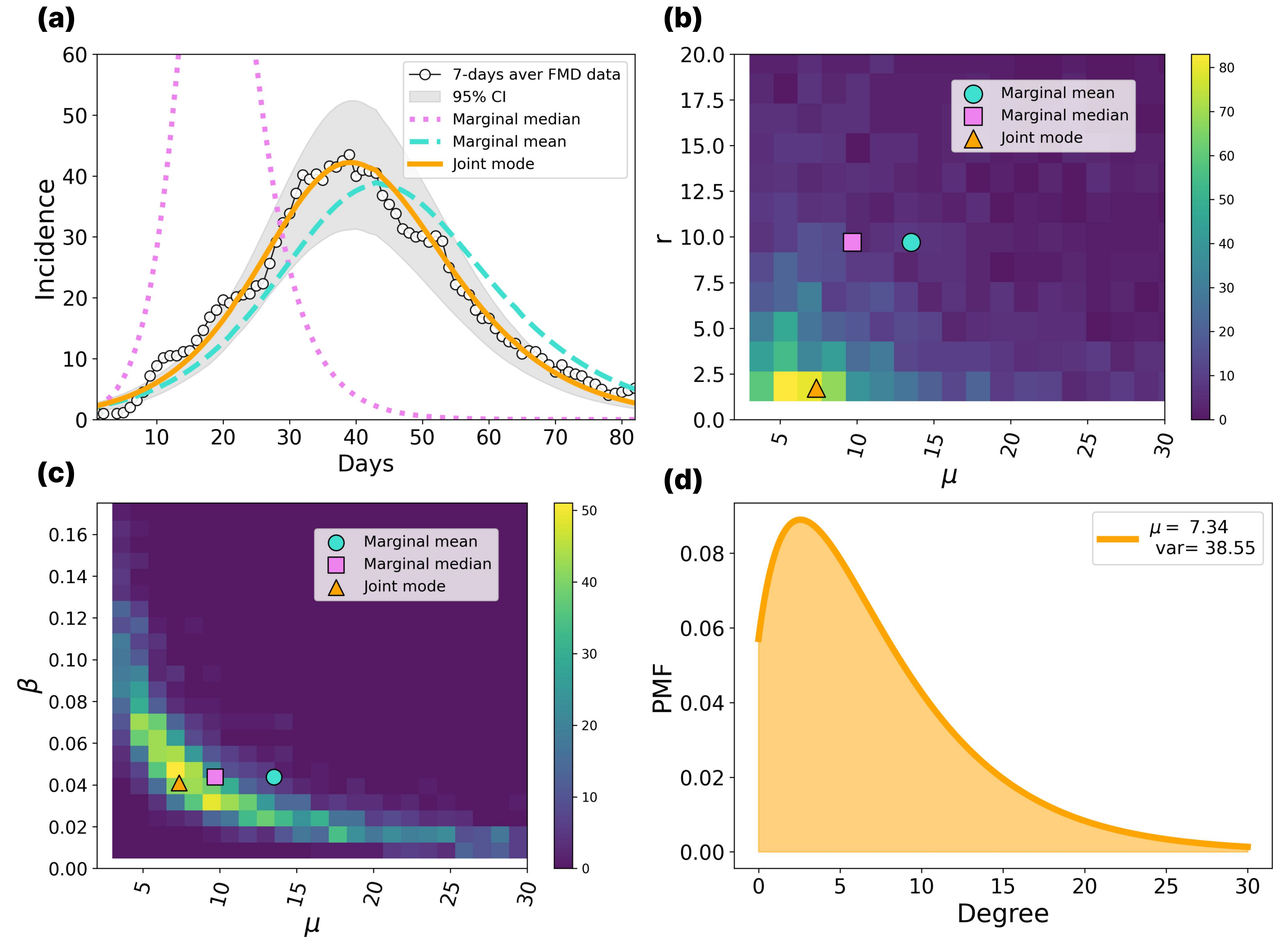}
	\caption{\textbf{Summary of inference results for the 2001 Foot-and-Mouth Disease outbreak in the UK.}  
(a) Evolution of the incidence based on parameter estimates from the three methods, alongside the 95\% credible interval from the posterior distribution. The observed FMD incidence, smoothed using a 7-day moving average, is also shown for comparison.  
(b) and (c) Heatmaps of the posterior distribution projections, showing parameter density across the $(\beta, \mu)$ and $(r, \mu)$ spaces, respectively. Each projection includes point estimates obtained using the \textit{marginal mean}, \textit{marginal median}, and \textit{joint mode}.  
(d) Inferred degree distribution obtained using the \textit{joint mode} estimation method.  }
	\label{fig:FMD_1}
\end{figure}

We generate five independent Markov chains, each running for 100,000 iterations, following the methodology outlined in Section~\ref{sec:inf_methods}. To construct the posterior distribution, we discard the first 50,000 iterations as burn-in and thin the remaining samples by selecting every 50th iteration to mitigate autocorrelation. From the resulting posterior, we estimate the five parameters $(\beta, \gamma, \mu, r, \rho)$ using three different methods: \textit{marginal mean}, \textit{marginal median}, and \textit{joint mode}, as shown in \ref{fig:diagram}. The \textit{joint mode} is computed using the mean-shift algorithm, which approximates the kernel density based on the posterior sample.  

In Fig.~\ref{fig:FMD_1}(a), we compare the EBCM solutions with real epidemic data using different parameter estimation methods. As shown, the parameters obtained via the \textit{marginal mean} fail to capture the system's behaviour. While the EBCM integrated with parameters estimated through the \textit{marginal median} provides a better fit, it still does not fully capture the original data. However, the EBCM solution obtained using parameters estimated via the \textit{joint mode} exhibits excellent agreement with the epidemic data. Additionally, we sample a subset of parameters, $\mathbf{S}$, from the posterior distribution to construct a 95\% credible interval for the daily new infections, as mentioned in the bottom right of \ref{fig:diagram}. Notably, the prediction based on the \textit{joint mode} lies at the centre of this interval. This is further illustrated in Fig.~\ref{fig:FMD_1}(b)-(c), where we present projections of the full posterior distribution for the $(\beta, \mu)$ and $(r, \mu)$ parameter spaces, respectively. These figures highlight that the \textit{marginal mean} and \textit{marginal median} estimates lie outside the region of highest posterior density. On the other hand, point estimates obtained using the \textit{joint mode} are closer to the densest regions of the posterior projections when compared to the estimates from the marginal distributions. The discrepancy is likely due to the presence of long tails in the posterior distributions, which obscure important correlations between parameters and lead to suboptimal point estimates. In Table~\ref{table:FMD_table_1}, we provide the point estimates for each method, along with the corresponding values of the $MSE$ calculated from the infeed incidence.

\begin{table}[b!]
    \centering
    \begin{tabular}{ | l  l  l l |}
    \hline
    Parameter & \textit{marginal mean} & \textit{marginal median} & \textit{joint mode} \\ \hline
    $\beta$ & 0.043 & 0.036 & 0.041\\ \hline
    $\gamma$ & 0.30 & 0.28 & 0.32\\ \hline
    $\mu$ & 13.51 & 9.70 & 7.34\\ \hline
    $r$ & 9.71 & 5.25 & 1.72\\ \hline
    Variance & 32.32 & 27.58 & 38.55\\ \hline
    $R_0$ & 1.89 & 1.32& 1.31\\ \hline \hline
    MSE Incidence & 3.44 & 1.03 & 0.76\\ \hline
    \end{tabular}
    \caption{Estimated parameter values obtained using three inference methods: \textit{marginal mean}, \textit{marginal median}, and \textit{joint mode}, for the the Foot-and-Mouth disease outbreak. The table also reports the Mean Squared Error (MSE) between the predicted incidence evolution and the observed data.}
    \label{table:FMD_table_1}
\end{table}


Additionally, in Fig.~\ref{fig:FMD_1}(d), we show the degree distribution with values of the mean and variance that are in line with expectations since highly connected markets were disproportionately affected at the beginning of the outbreak leading to a marked reduction in network heterogeneity \cite{davies2002foot}.
While the model successfully captures the epidemic curve, the inferred contact network structure cannot be directly validated against ground truth data, as no empirical contact network is available for comparison.

\begin{figure}[t!]
	\centering
	\includegraphics[width=0.5\textwidth]{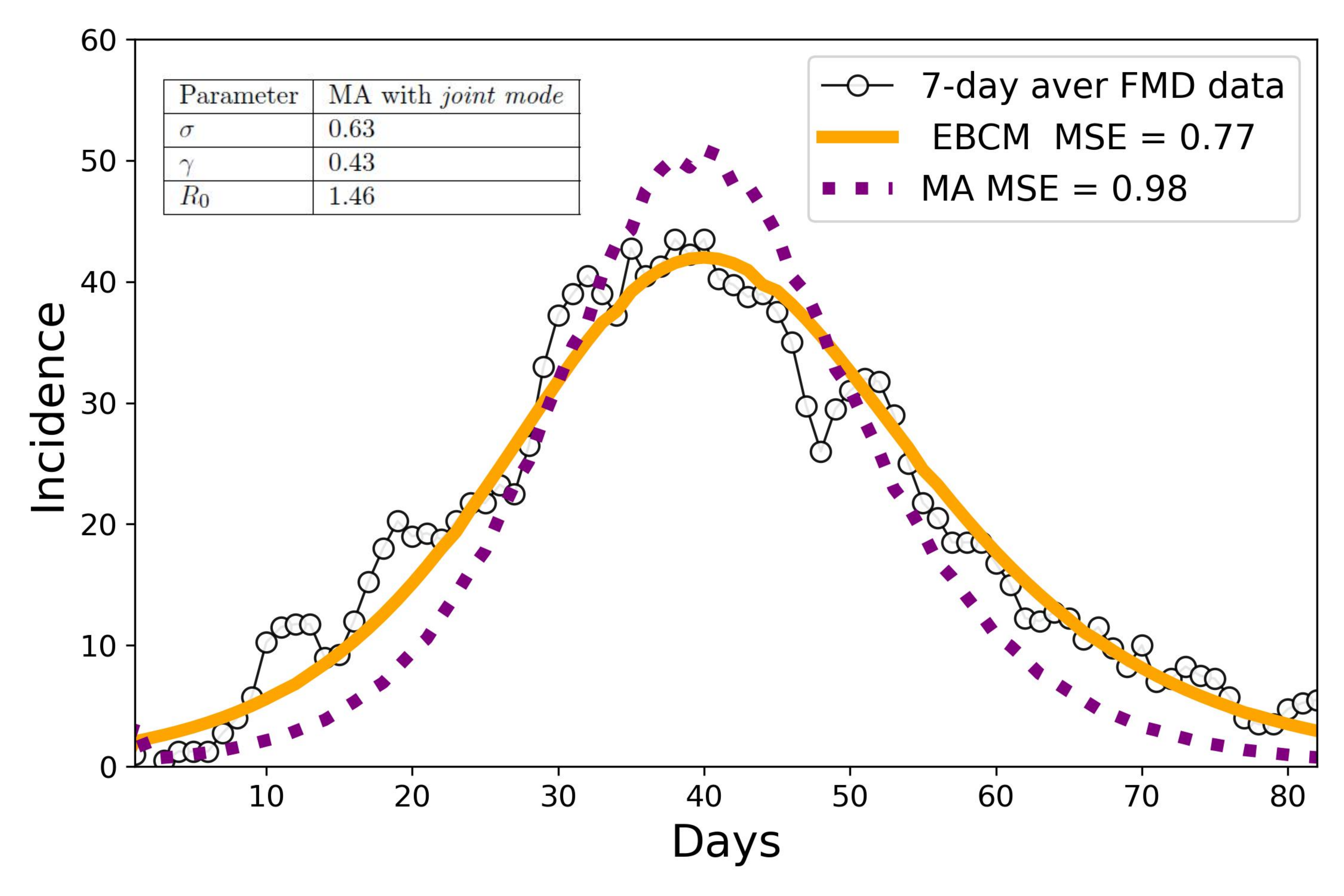}
	\caption{\textbf{Comparison of predicted incidence of infections using the EBCM and mass-action model for Foot-and-Mouth Disease (FMD).} The estimated incidence curves are presented for the DSA framework using the Edge-Based Compartmental Model (EBCM, shown as the orange curve) and the Mass-Action (MA) model (depicted as purple squares). Point estimates were derived using the joint mode for both models. The table in the top-left corner displays the results of the MA model, while the EBCM estimates can be found in Table \ref{table:FMD_table_1}. The black circled line represents the 7-day moving average of observed FMD cases.}
	\label{fig:EBCM_MA}
\end{figure}

Furthermore, we compare the predictions made by the EBCM, with the ones made by using the MA model. In this case, the parameters exhibit a lower correlation, resulting in a posterior distribution without long tails. Consequently, the \textit{marginal median}, \textit{marginal mean}, and \textit{joint mode} produce similar estimates. In the table at the top-left corner in Figure ~\ref{fig:EBCM_MA}, we show the estimates of $\beta$, $\gamma$, and $R_0$. Using these estimates, we can compare the MA and EBCM models in their ability to describe the original data. In Fig. ~\ref{fig:EBCM_MA}, we plot the incidence along with the MSE calculated for each case.
The EBCM, which accounts for an explicit contact structure, provides a closer fit to the original data compared to the classical MA model. This is clearly evidenced by the lower MSE values achieved by the EBCM, suggesting that FMD was more likely to have spread through an explicit contact structure.

\subsection{COVID-19 data from Seoul, South Korea }
\label{subsubsec:Korea}

In this section, we extend our analysis to the first wave of the COVID-19 pandemic in Seoul, South Korea, covering an 84-day period from January 26 to April 18, 2020 \cite{Seoul_Covid}. This dataset provides information on the time of symptom onset for each individual, as well as the date of their positive test result.  
For our analysis, we assume that the time of infection coincides with the onset of symptoms. Given the strict isolation measures in place, we treat the time of a positive test result as the effective recovery time, as individuals were promptly isolated upon testing positive. Using this dataset, we can aggregate the information to track the evolution of incidence (the number of new infections per day, as in Section \ref{subsubsec:FMD}), prevalence (the total number of currently infected individuals at any given time), and daily recoveries (the number of individuals recovering each day).

Notably, the Seoul COVID-19 dataset also includes contact pattern data, recording the number of contacts each infected individual had between symptom onset and recovery. This allows us to use the contact data as ground truth for evaluating the network structure characteristics inferred by the EBCM. By comparing the inferred network properties with empirical contact data, we assess the capability of the model to recover meaningful structural information from epidemic observations. To perform our analysis, we consider the Negative Binomial degree distributions and follow the same procedure as in Section \ref{subsubsec:FMD}. Additionally, we consider the likelihood $\ell_1+\ell_2$, which is able to accommodate the most complete type of data.

\begin{figure}[t!]
	\centering
	\includegraphics[width=0.75\textwidth]{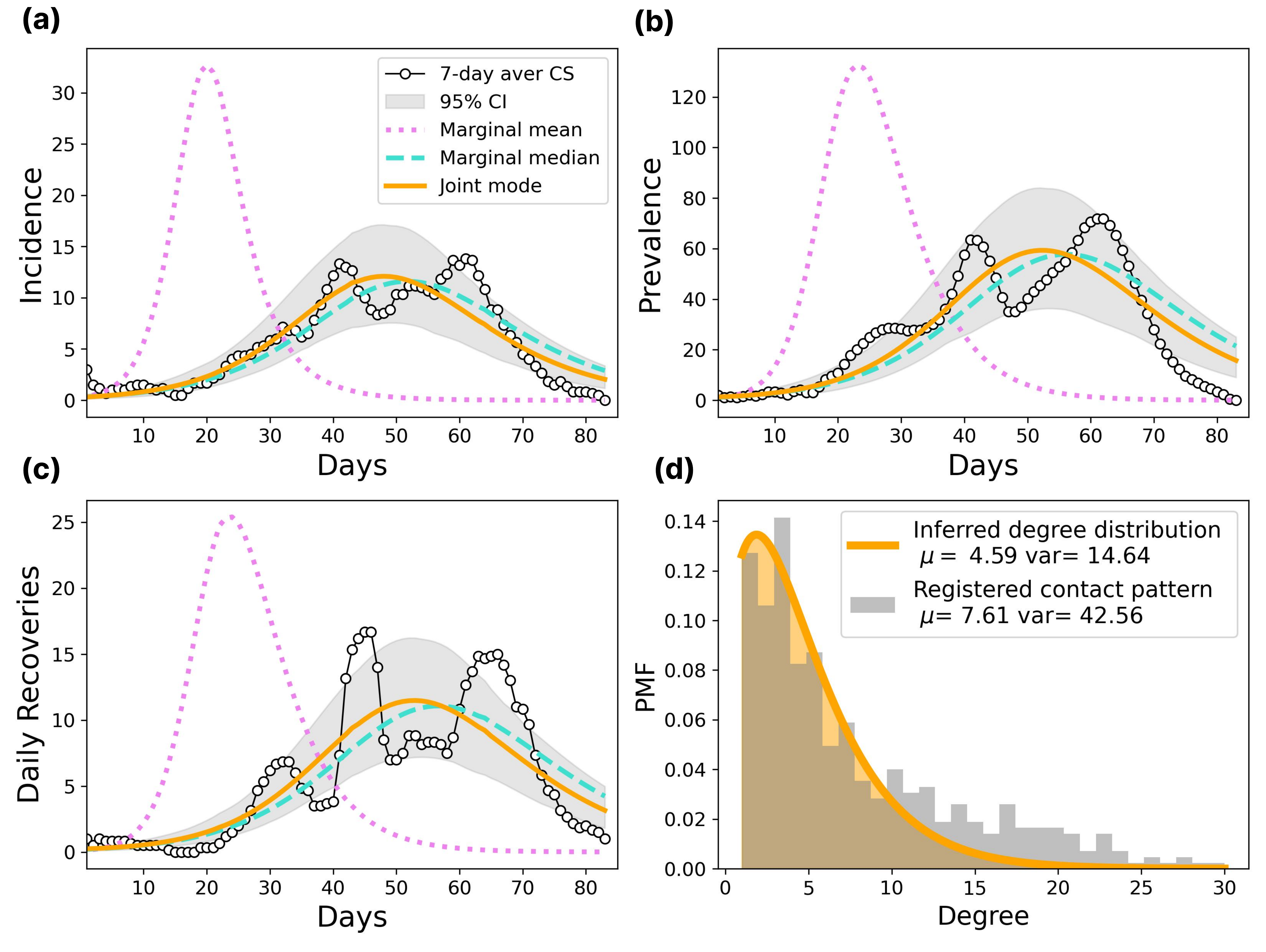}
	\caption{\textbf{Inference results for the first wave of COVID-19 in Seoul.}  
Panels (a), (b), and (c) depict the time evolution of prevalence, incidence, and daily recoveries, respectively, estimated using three inference methods: \textit{marginal mean} (pink dotted line), \textit{marginal median} (light blue dashed line), and \textit{joint mode} (orange solid line). These estimates are compared to the 7-day moving average of observed data from the first wave of COVID-19 in Seoul (CS), represented by black empty circles. Panel (d) presents the inferred probability mass function derived from epidemic data, alongside the recorded contact data. }
	\label{fig:Korean_1}
\end{figure}

In Fig. ~\ref{fig:Korean_1}, we show the predictions based on the EBCM for prevalence, incidence, and daily recoveries.

As with the FMD disease dataset (Section \ref{subsubsec:FMD}), the set of parameters estimated using the \textit{joint mode} provides a temporal evolution of the system that aligns closely with the original data. In this case, the \textit{marginal median} also performs well, yielding results that are closer to those obtained with the \textit{joint mode}. In contrast, the \textit{marginal mean} produces the least accurate results, failing to capture the behaviour of the system. These findings are illustrated in Fig.~\ref{fig:Korean_1}(a), \ref{fig:Korean_1}(b), and \ref{fig:Korean_1}(c), which show the evolution of incidence, prevalence, and daily recoveries for each method. Additional evidence for these results is provided by the Mean Squared Error (MSE) values in Table~\ref{table:Table_Korean}.

\begin{table}[b!]
    \centering
    \begin{tabular}{ | l  l  l  l |}
    \hline
    Parameter & \textit{marginal mean} & \textit{marginal median} &  \textit{ joint mode}\\ \hline
    $\beta$ & 0.042 & 0.035 & 0.052\\ \hline
    $\gamma$ & 0.192 & 0.192 & 0.193\\ \hline
    $\mu$ & 12.80 & 8.96 & 4.59\\ \hline
    Variance & 16.82 & 12.30& 14.64\\ \hline
    $R_0$ & 2.35  & 1.43 & 1.44\\ \hline \hline
    MSE Incidence & 1.14 & 0.48 & 0.41\\ \hline
    MSE Prevalence & 4.65 & 2.65 & 2.37 \\\hline 
    MSE Daily Recovered & 1.07 & 0.65& 0.6\\ \hline
    \end{tabular}
    \caption{Estimated parameter values obtained using three inference methods: \textit{marginal mean}, \textit{marginal median}, and \textit{joint mode}, for the 7-day moving average of the first wave of COVID-19 in Seoul (CS). The table also reports the Mean Squared Error (MSE) between the predicted epidemic trajectory and the observed data.}
    \label{table:Table_Korean}
\end{table}

In Fig.~\ref{fig:Korean_1}(d), we compare the inferred degree distribution with the one obtained from the contact data. While the inferred distribution captures key characteristics of the empirical data, the point estimates derived from the \textit{joint mode} yield a lower average degree and variance than those observed in the real network. Notably, the Negative Binomial distribution struggles to fully reproduce the long tail present in the empirical contact data, suggesting that higher-degree individuals may be under-represented in the inferred network structure.

\begin{figure}[t!]
	\centering
	\includegraphics[width=0.85\textwidth]{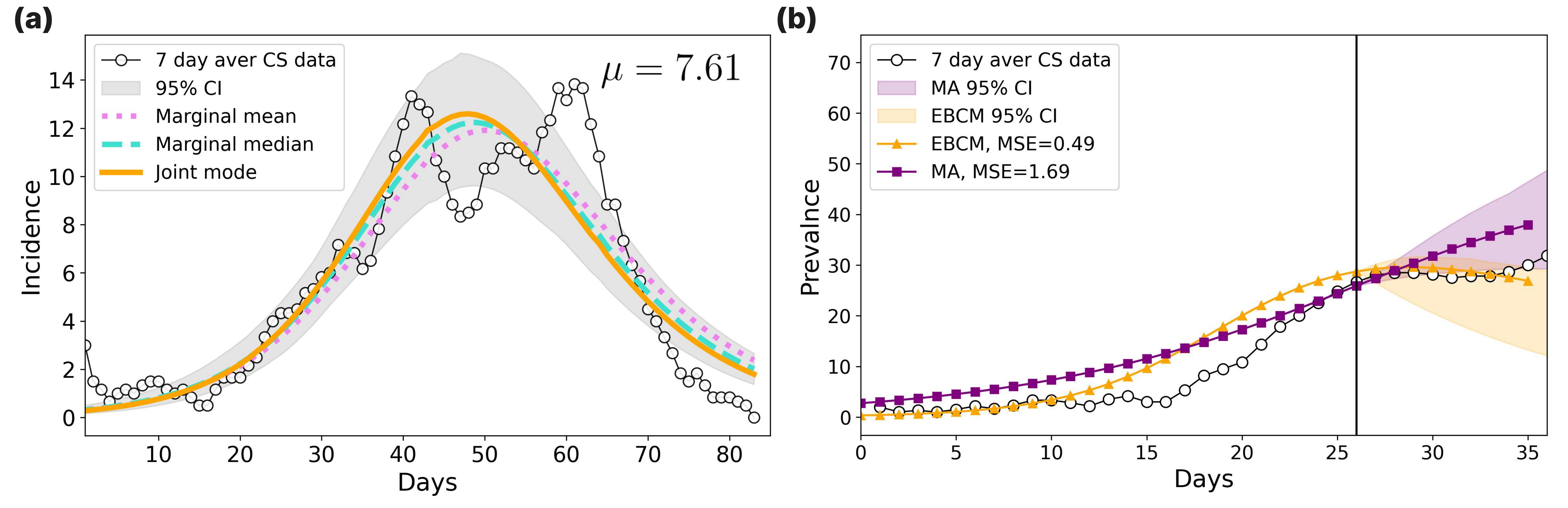}
	\caption{\textbf{Additional analysis of the first wave of COVID-19 in Seoul:} Panel (a) shows the evolution of the incidence obtained from the point estimates using \textit{marginal mean}, \textit{marginal median}, and \textit{joint mode} for the case where the average degree is known and fixed at $\mu=7.61$. Panel (b) shows the results obtained using partial data. Specifically, the first 26 days of data were used to forecast the following 10 days. In the figure, we show the prediction obtained using the EBCM in orange triangles, MA in square purple, and the 7 days moving average of the COVID-19 in Seoul (CS) in the black circles. }
	\label{fig:fig8}
\end{figure}

Furthermore, we can use the contact data to fix the average degree. In this case, we fix  $\mu=7.61$, which is the average number of contacts entered in the survey. We can apply the inference procedure to find $\beta$,  $\gamma$, and the parameter $r$ of the Negative Binomial distribution. Table ~\ref{table:Fixed} presents the point estimates obtained from the three methods: \textit{marginal mean}, \textit{marginal median}, and \textit{joint mode}. In this scenario, the three methods produce very similar estimates. Furthermore, in Fig.~\ref{fig:fig8}(a), we observe the incidence predicted by the three methods, which are nearly identical. This indicates that fixing the average degree results in a much better-behaved posterior distribution. It is important to note that the contact data collected should be interpreted with caution. This data is highly dependent on an individual's perceptions and, as such, cannot be treated as a definitive ground truth for contact patterns.


\begin{table}[b!]
    \centering
    \begin{tabular}{ | l  l  l  l |}
    \hline
    Parameter & \textit{marginal mean} & \textit{marginal median} & \textit{joint mode}\\ \hline
    $\beta$ & 0.044 & 0.044 & 0.043\\ \hline
    $\gamma$ & 0.196 & 0.196 & 0.193\\ \hline
    $Variance$ & 8.77& 9.33 & 12.10\\ \hline
    $R_0$ & 1.42  & 1.44 & 1.47\\ \hline \hline
    $MSE$   Incidence & 0.45 & 0.42 & 0.41\\ \hline
    $MSE$ Prevalence & 2.44 & 2.30 & 2.25\\ \hline
    $MSE$   New Recovered & 0.63 & 0.61 & 0.60\\ \hline
    \end{tabular}
    \caption{Estimated parameter values obtained using three inference methods: \textit{marginal mean}, \textit{marginal median}, and \textit{joint mode}, for the 7-day moving average of the first wave of COVID-19 in Seoul (CS) for a fixed average degree $\mu=7.61$, obtained from the COVID-19 Survey data. The table also reports the Mean Squared Error (MSE) between the predicted epidemic trajectory and the observed data}
    \label{table:Fixed}
\end{table}

Finally, to further demonstrate the potential of this framework, we performed parameter estimation—including the mean degree—using only partial data, specifically up to a cut-off time \( T_c \) before the 84th day. In this case, we used the first 26 days of data to infer parameters and then forecasted the number of new infections for the following 10 days. The 95\% credible interval was obtained by sampling from the full posterior distribution, while point estimates were determined using the \textit{joint mode}. 
For comparison, we performed the same analysis using the mass-action (MA) model. The results are shown in Fig.~\ref{fig:fig8}(b). The forecasted incidence was compared to the observed data using the mean squared error (MSE). The results indicate that the EBCM provides the most accurate forecast.

\section{Discussion}
\label{sec:Discussion}

Understanding the contact patterns of individuals during an epidemic remains a fundamental challenge in infectious disease modelling. Inferring the underlying network structure of an epidemic process is particularly difficult, even when the goal is not to reconstruct the entire network but rather to estimate key characteristics, such as the average degree or variance. A major obstacle in this inference process is the practical identifiability between connectivity and infectivity, a challenge previously highlighted in the literature \cite{britton_bayesian_2002,kiss_towards_2023}.

In this study, we introduced a framework for inferring network properties from epidemic data by integrating the Dynamic Survival Analysis (DSA) framework with the Edge-Based Compartmental Model (EBCM). The EBCM provides a compact yet effective representation of the epidemic process, where the degree distribution is incorporated as a model parameter. By combining this with the flexibility of DSA, our approach enables the inference of crucial properties of the contact network that drive epidemic spreading.

We validated this framework using synthetic epidemic data generated via Gillespie simulations, considering networks with Poisson and Negative Binomial degree distributions. By applying the DSA-EBCM approach, we sampled from the joint posterior distribution of both disease and network parameters. Despite the inherent correlation between the infection rate, the recovery rate, and network properties such as the average degree, the posterior distributions consistently concentrated around the true parameter values. This allowed for an accurate reconstruction of both the epidemic dynamics and key network characteristics, despite relying solely on epidemic time-series data in which the network structure is only implicitly present.

Beyond synthetic data, we tested our methodology on two real-world outbreaks: the 2001 Foot-and-Mouth Disease (FMD) epidemic in the UK and the first wave of COVID-19 in Seoul, South Korea. In both cases, the framework successfully produced robust posterior distributions despite correlations between parameters. For the FMD dataset, we observed a multidimensional posterior distribution with long tails and a strong inverse correlation between $\beta$ and $\mu$. Nevertheless, as in the synthetic cases, a high-density region in parameter space provided the best description of the original data. Notably, we found that the \textit{joint mode}, which represents the point of highest density in the sample of the posterior distribution, although more computationally demanding and less exact, yielded better point estimates compared to the \textit{marginal median} or \textit{marginal mean}, as it minimized the mean squared error. 

For the Seoul COVID-19 dataset, we observed similar results, with the \textit{joint mode} again emerging as the most accurate estimator of the epidemic process. Additionally, our inferred average degree closely matched independent contact data collected during the outbreak, reinforcing the validity of the approach. The framework also demonstrated its predictive capabilities by generating a short-term forecast with a 95\% credible interval for the epidemic's progression over a 10-day period.

This study demonstrates that meaningful insights about underlying contact structures can be extracted solely from epidemic data without requiring explicit network observations. Future work could explore the extent to which this method can distinguish between homogeneous and heterogeneous network structures based only on outbreak dynamics. Furthermore, the approach could be extended to other spreading processes, such as information diffusion, where higher-order interactions may play a significant role. Comparing inferred network properties across different geographic regions—such as cities, counties, or states—could also provide insights into the diverse mechanisms that shape disease transmission.

Overall, our findings highlight the potential of integrating the EBCM and other network-based mean-field models with DSA to infer hidden contact structures from limited epidemic data. This approach provides a powerful tool for reconstructing essential network characteristics, improving epidemic forecasting, and enhancing our understanding of infectious disease spread in real-world settings.

\label{sec:discussion}
\section*{Acknowledgements}
A.G. acknowledges the PhD studentship support from Northeastern University London. A.G. and I.Z.K. acknowledge useful discussion with Mauricio Santillana.
B.C. and G.H.P. acknowledge the Basic Science Research Program through the NRF funded by the Ministry of Education (RS-202300245056). B.C acknowledges a grant of the project The Government-wide R\&D to Advance Infectious Disease Prevention and Control (HG23C1629).

\section*{Data availability statement}
Code and synthetic datasets generated and analyzed during the current study are available from the corresponding author upon reasonable request.

\bibliography{project1}
\bibliographystyle{unsrt}

\pagebreak

\renewcommand{\figurename}{Supplementary Figure}
\renewcommand{\tablename}{Supplementary Table}

\setcounter{figure}{0}
\setcounter{table}{0}

\section*{Supplementary Material}
This Supplementary Material provides a detailed and systematic analysis of the performance of DSA when used with synthetic data. In particular, we focus on understanding how the goodness of estimation depends on factors such as (i) match/mismatch between networks used to generate the data versus network used in the EBCM, (ii) type of data ($l_1$, $l_1+l_2$,$l_1+l_3$,$l_3$), and (iii) various combinations of fixing some and estimating other parameters. The SM is divided into subsections focusing on: the estimation of $R_0$, estimation of $\gamma$, estimation of $\beta$, estimation of $\mu$, and finally, a consideration of the goodness of fit on all the parameters mentioned above when using partial data. 

\label{Append:Sim}


\subsection*{Decoupling network density and rate of infection}
\label{Append:Dist}
One approach to addressing identifiability is to restrict the inference by fixing one of the parameters to be estimated. This method helps in understanding how the inference depends on which parameter is known or inferred. To test this, we repeated the inference procedure across all scenarios and data combinations, while fixing $\beta=0.2$.

Here, we focus on two general behaviors observed when fixing epidemic parameters. In the first column of Supplementary Fig.~\ref{fig:Sim_study_fixed}, we show the case where Negative Binomial (NB) data is fitted with the NB model, with the restriction of fixing $\beta = 0.2$. In this scenario, all estimates for $R_0$, $\gamma$, and $\mu$ are accurate relative to the original parameters and exhibit much smaller variance compared to the unrestricted case shown in the Figure 2 of the main text. This illustrates a common outcome: fixing one parameter improves the accuracy and reduces the variance of the remaining parameter estimates. However, in some cases, fixing a parameter does not yield better results. The second column of Supplementary Fig.~\ref{fig:Sim_study_fixed} shows the results for Negative Binomial data fitted with a Poisson model, with $\beta$ fixed at 0.2. In this case, the accuracy of the estimates for $\gamma$ and $\mu$ is worse than in the unrestricted scenario, with a noticeable increase in bias for these parameters. In contrast, $R_0$ is estimated more accurately. This occurs because the estimation procedure prioritizes fitting the observed data as closely as possible. When one parameter is fixed, the remaining parameters adjust to compensate, leading to more accurate $R_0$ estimates but introducing significant bias in $\gamma$ and $\mu$.

\begin{figure}[h!]
	\centering
	\includegraphics[width=0.8\textwidth]{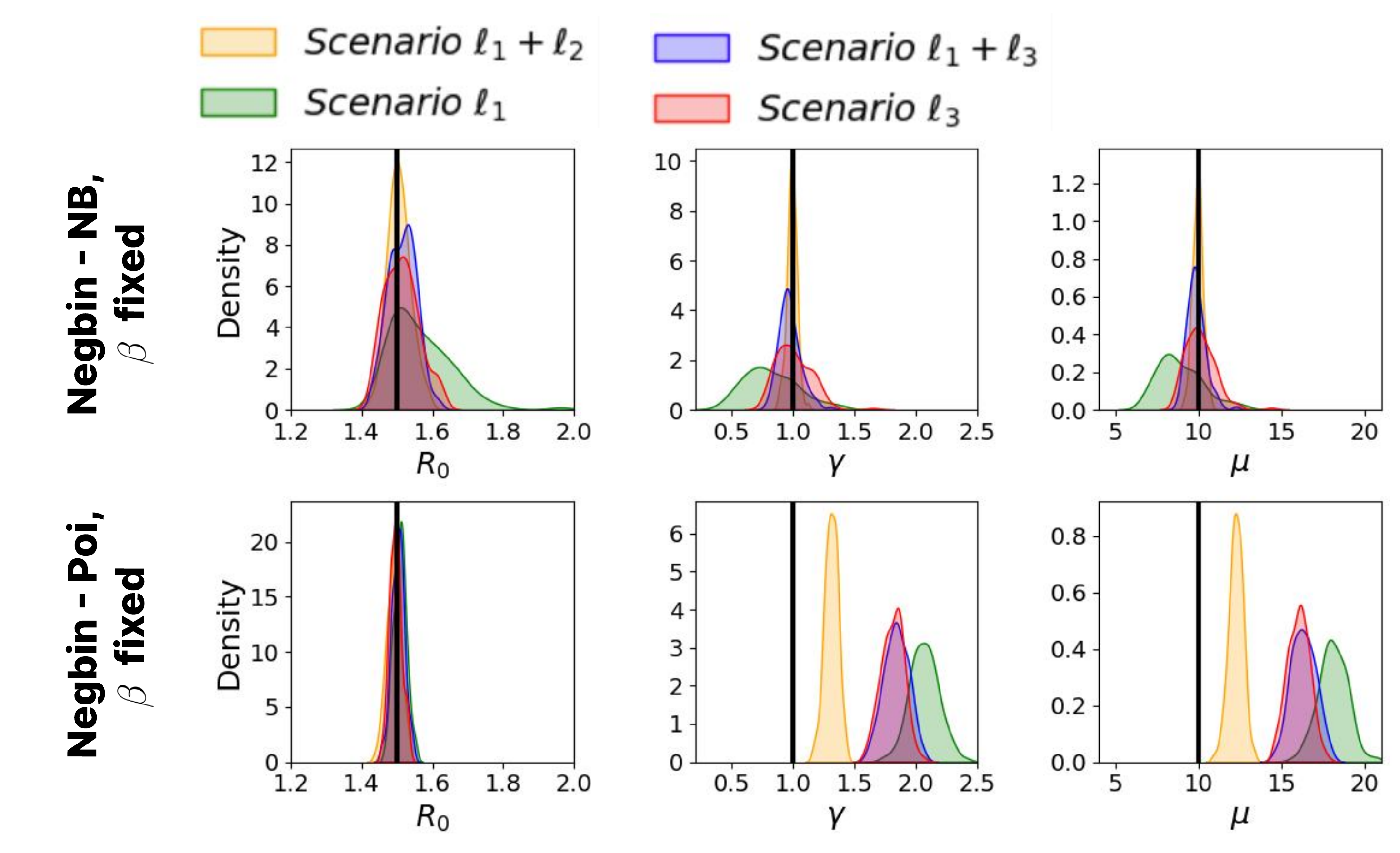}
	\caption{\textbf{Distributions of the \textit{marginal medians} for model parameters $\mu$ and $\gamma$ and the basic reproduction number $R_0$ inferred from the DSA framework, for a fixed value of $\beta$ of 0.2.} 
    Results are obtained by fitting 200 distinct realizations of Gillespie simulations with parameters $\beta = 0.2$, $\gamma = 1$. Networks exhibiting a Negative Binomial (Negbin) degree distribution were generated using parameters $\mu = 10$, $r = 1$, and $p = 0.091$, while networks with a Poisson degree distribution were generated with $\mu = 10$. All networks have $N=10000$ nodes and $\rho=10^{-4}$.}
	\label{fig:Sim_study_fixed}
\end{figure}

\subsection*{Estimation of $R_0$}
\label{Append:R0}
\begin{figure}[h!]    
\centering
\includegraphics[width=0.80\linewidth]{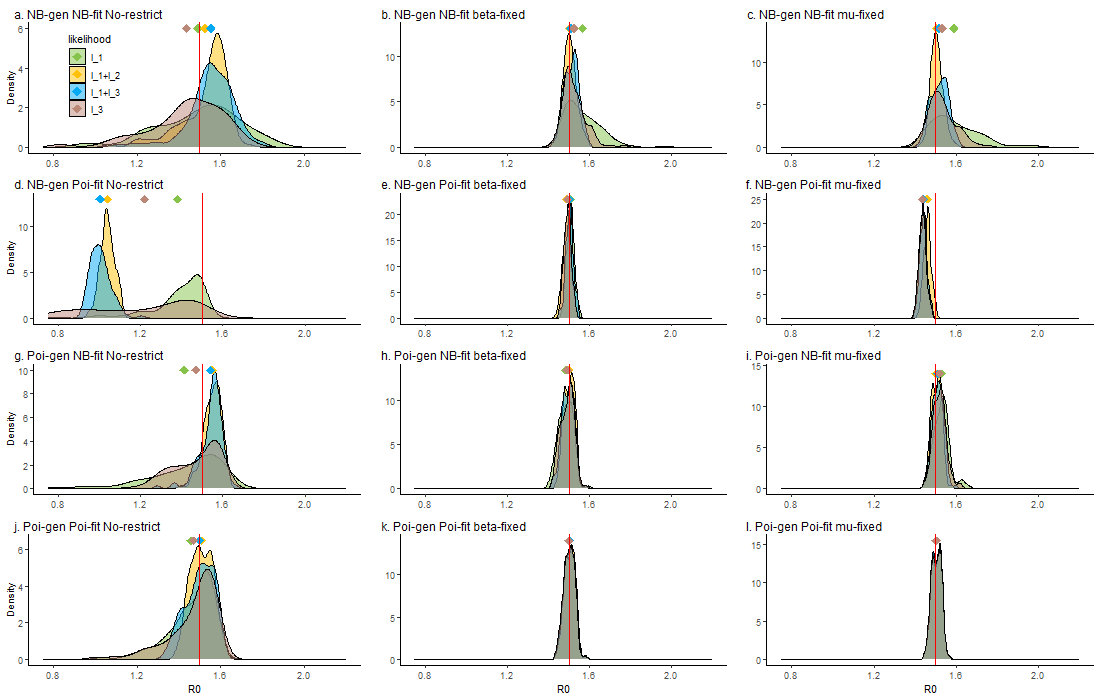}
	\caption{\textbf{Fitting summary of $R_0$.} Each panel depicts the density of 100 posterior medians under the four scenarios. The simulation data used the negative binomial degree distribution for the upper two rows and the Poisson degree distribution for the lower two rows. The first and fourth rows are correctly matched between the data and model; the first is for negative binomial, and the fourth is for Poisson, respectively. the second and third rows are mismatched between the data and the model. The first column panels denoted the result when we estimated all parameters. The second column panels denoted the result when we fixed $\beta$, and the third column is the result with $\mu$ fixed.}
	\label{fig:R0}
\end{figure}

Supplementary Fig. \ref{fig:R0} and Supplementary Table \ref{tb:R0} denote the simulation summary of the basic reproduction number $R_0$. Each panel shows a density plot of 100 posterior medians according to the four scenarios. The first two rows represent the fitting results with Negative Binomial degree distribution data, and the third and fourth rows are that of Poisson degree distribution data. The first row represents the result for the correctly matched NB model, and the fourth row for the Poi model. The second row is the result of the mismatched Poi model, and the third is the mismatched NB model. The first column is the result when we estimate all parameters without any constraints. The second and third columns are the results of estimating only the remaining parameters with $\beta$ and $\mu$ fixed at their true values of 0.2 and 10, respectively.

Thus, panel (a) displays the result achieved when the data and model are matched with the same negative binomial distribution and all parameters are estimated simultaneously using the MCMC method. The average of the 100 estimates shows accurate results, ranging from  1.44 to 1.55 without a significant difference from the true value of 1.5 (refer to the first row and the first four columns in  Supplementary Table \ref{tb:R0}). As shown in the density plot, scenario $l_1+l_2$  and scenario $l_1+l_3$, which use relatively more information, have smaller MSEs of 0.0145 and 0.0164, respectively, than scenarios $l_1$ and $l_3$. Our fitting method gives an accurate estimate of $R_0$ if we fit the model using the correct degree distribution. As discussed in Section \ref{sec:Results}, there is an identifiability problem between the infection rate $\beta$ and the mean degree $\mu$ of the degree distribution. To check the effect of the identifiability between $\beta$ and $\mu$ in the estimation, we fix one of the two parameters at their true values of 0.2 or 10, then estimate the other parameters alone. Panels (b) and (c) show that the fitting results are a little improved. All the MSEs are also improved for all scenarios (the last eight columns and the first row in Supplementary Table \ref{tb:R0}. Next, let us look at the results when the data and model are mismatched. In panel (d), we can see serious bias for scenarios $l_1+l_2$  and scenario $l_1+l_3$ and variance for scenarios $l_1$ and $l_3$. These inaccuracies are dramatically reduced when we fixed $\beta$ (panel (e)) or fixed $\mu$ (panel (e). However, there is a slight bias when we fixed $\mu$, but the difference is insignificant. 
\begin{table}[b!]
    \centering
    \resizebox{0.8\textwidth}{!}{ 
    \begin{tabular}{c|c|c|r|r|r|r|r|r|r|r|r|r|r|r}
        \hline
        \multirow{2}{*}{\begin{tabular}{c}Degree \\ distribution \end{tabular}} & \multirow{2}{*}{\begin{tabular}{c}Fitting \\ model \end{tabular}}  & \multirow{2}{*}{Statistics} & \multicolumn{4}{c|}{No restriction} & \multicolumn{4}{c|}{$\beta$ fixed} & \multicolumn{4}{c}{$\mu$ fixed} \\ \cline{4-15}
        & & & \multicolumn{1}{c|}{$l_1$} & \multicolumn{1}{c|}{$l_1+l_2$} & \multicolumn{1}{c|}{$l_1+l_3$} & \multicolumn{1}{c|}{$l_3$} & \multicolumn{1}{c|}{$l_1$} & \multicolumn{1}{c|}{$l_1+l_2$} & \multicolumn{1}{c|}{$l_1+l_3$} & \multicolumn{1}{c|}{$l_3$} & \multicolumn{1}{c|}{$l_1$} & \multicolumn{1}{c|}{$l_1+l_2$} & \multicolumn{1}{c|}{$l_1+l_3$} &  \multicolumn{1}{c}{$l_3$} \\ \hline
        \multirow{8}{*}{\begin{tabular}{c}Negative \\ binomial \\ degree \\ distribution\end{tabular}} 
        & \multirow{4}{*}{\begin{tabular}{c}NB \\ model \end{tabular}}
        & Mean & 1.4916 & 1.5242 & 1.5531 & 1.4369 & 1.5665 & 1.5046 & 1.5120 & 1.5251 & 1.5906 & 1.5061 & 1.5183 & 1.5331 \\ \cline{3-15}
        &  & MSE & 0.0394 & 0.0145 & 0.0164 & 0.0367 & 0.0121 & 0.0010 & 0.0014 & 0.0039 & 0.0216 & 0.0009 & 0.0021 & 0.0057 \\ \cline{3-15}
        &  & Bias$^2$ & 0.0001 & 0.0006 & 0.0028 & 0.0040 & 0.0044 & 0.0000 & 0.0001 & 0.0006 & 0.0082 & 0.0000 & 0.0003 & 0.0011 \\ \cline{3-15}
        &  & Var & 0.0397 & 0.0141 & 0.0137 & 0.0330 & 0.0078 & 0.0010 & 0.0013 & 0.0033 & 0.0135 & 0.0009 & 0.0018 & 0.0046 \\ \cline{2-15}
       & \multirow{4}{*}{\begin{tabular}{c}Poi \\ model \end{tabular}}
        & Mean & 1.3808 & 1.0389 & 1.0042 & 1.2200 & 1.5076 & 1.4894 & 1.5029 & 1.4932 & 1.4367 & 1.4605 & 1.4408 & 1.4366 \\ \cline{3-15}
        &  & MSE & 0.0381 & 0.2138 & 0.2483 & 0.1371 & 0.0004 & 0.0005 & 0.0003 & 0.0003 & 0.0043 & 0.0019 & 0.0038 & 0.0043 \\ \cline{3-15}
        &  & Bias$^2$ & 0.0142 & 0.2126 & 0.2458 & 0.0784 & 0.0001 & 0.0001 & 0.0000 & 0.0000 & 0.0040 & 0.0016 & 0.0035 & 0.0040 \\ \cline{3-15}
        &  & Var & 0.0241 & 0.0012 & 0.0025 & 0.0593 & 0.0003 & 0.0004 & 0.0003 & 0.0003 & 0.0003 & 0.0003 & 0.0003 & 0.0003 \\ \hline
        
        \multirow{8}{*}{\begin{tabular}{c}Poisson \\ degree \\ distribution\end{tabular}} 
        & \multirow{4}{*}{\begin{tabular}{c}NB \\ model \end{tabular}}
        & Mean & 1.4159 & 1.5492 & 1.5414 & 1.4709 & 1.4834 & 1.5010 & 1.4945 & 1.4918 & 1.5304 & 1.5042 & 1.5118 & 1.5194 \\ \cline{3-15}
        &  & MSE & 0.0436 & 0.0043 & 0.0051 & 0.0138 & 0.0015 & 0.0007 & 0.0009 & 0.0011 & 0.0024 & 0.0006 & 0.0009 & 0.0014 \\ \cline{3-15}
        &  & Bias$^2$ & 0.0071 & 0.0024 & 0.0017 & 0.0008 & 0.0003 & 0.0000 & 0.0000 & 0.0001 & 0.0009 & 0.0000 & 0.0001 & 0.0004 \\ \cline{3-15}
        &  & Var & 0.0369 & 0.0019 & 0.0034 & 0.0131 & 0.0012 & 0.0007 & 0.0009 & 0.0010 & 0.0015 & 0.0006 & 0.0007 & 0.0010 \\ \cline{2-15}
        & \multirow{4}{*}{\begin{tabular}{c}Poi \\ model \end{tabular}}
        & Mean & 1.4624 & 1.5077 & 1.4991 & 1.4701 & 1.5032 & 1.5036 & 1.5029 & 1.5007 & 1.5041 & 1.5043 & 1.5044 & 1.5038 \\ \cline{3-15}
        &  & MSE & 0.0132 & 0.0029 & 0.0044 & 0.0153 & 0.0007 & 0.0007 & 0.0007 & 0.0007 & 0.0005 & 0.0005 & 0.0005 & 0.0005 \\ \cline{3-15}
        &  & Bias$^2$ & 0.0014 & 0.0001 & 0.0000 & 0.0009 & 0.0000 & 0.0000 & 0.0000 & 0.0000 & 0.0000 & 0.0000 & 0.0000 & 0.0000 \\ \cline{3-15}
        &  & Var & 0.0119 & 0.0028 & 0.0045 & 0.0145 & 0.0007 & 0.0006 & 0.0007 & 0.0007 & 0.0005 & 0.0005 & 0.0005 & 0.0005 \\ \hline
    \end{tabular}}
    \caption{\textbf{Summary statistics of $R_0$} This table summarizes the mean, MSE (mean squared error), squared bias, and variance of the 100 posterior medians of simulation according to the two fitting degree models and four scenarios. }
    \label{tb:R0}
\end{table}
Next, turn to the result for the data with the Poisson degree distribution. 
The third and fourth row in Supplementary Fig. \ref{fig:R0} reveals results of which data were generated from a Poisson degree distribution and model fitting using the NB model and Poi model, respectively. The Poi model fit consistently yields accurate and precise estimation across all four scenarios (Panel j). When the data and model are matched with a Poisson distribution, a consistency that is not surprising given the one less parameter to estimate compared to a negative binomial distribution.

Interestingly, the estimation results for mismatched cases between Poisson degree data and NB model fitting are also accurate (panel (g)). Comparing these results to those in panel (a) with the data generated by negative binomial degree distribution, we note that the bias is slightly larger, but the MSEs are smaller in the ninth row and first four columns in Supplementary Table \ref{tb:R0}, demonstrating an advantage of the NB model over the Poi model. Despite the NB model having one more parameter to estimate than the Poi model, the averages of the estimates of $p$ of the Negative Binomial distribution are consistently high, ranging from 0.98 to 0.99 according to the four scenarios.

These suggest that the negative binomial distribution converges to the Poisson distribution, and the estimation results of the NB and Poi models are remarkably similar. The estimation results with fixed $\beta$ or $\mu$ (panels (h)-(l)) show that both the Poi model and the NB model produced accurate and precise estimates.

\subsection*{Estimation of $\gamma$}
\label{Append:gamma}
As in Supplementary Fig. \ref{fig:R0}, Supplementary Fig. \ref{fig:ga} summarises the estimation results of $\gamma$. Supplementary Table \ref{tb:ga} also describes the summary statistics for estimating $\gamma$. In panel (a), we see that the estimation results, except for scenario $l_1$, are accurate. In particular, as scenario $l_1+l_2$ uses the infectious period as the data for estimation, which can directly reflect $\gamma$, the inverse of the mean of the infectious period, gives a completely accurate estimation. Even when a mismatched Poi model is applied, the estimation for scenario $l_1+l_2$ is also accurate (panel (d)). In fact, scenario $l_1 +l_2$ provided accurate and precise estimation results across the degree distributions and fitting models (refer to the second column in Supplementary Table \ref{tb:ga}.

\begin{figure}[t!]
	\centering
	\includegraphics[width=0.8\linewidth]{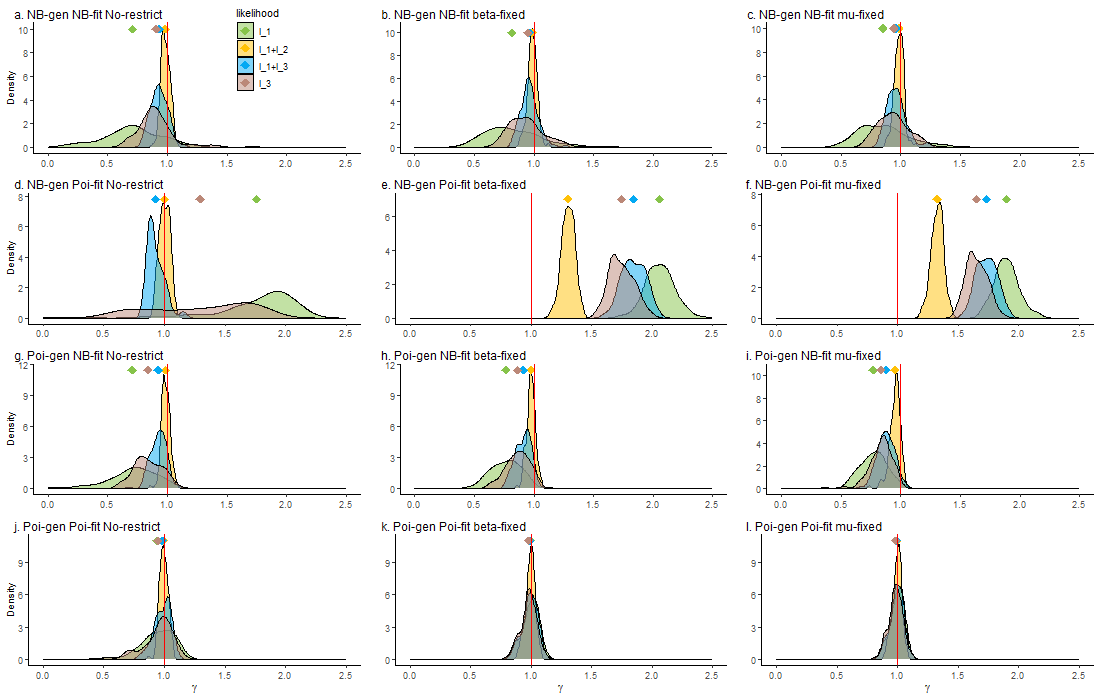}
	\caption{\textbf{Fitting summary of $\gamma$.} Each panel depicts the density of 100 posterior medians under the four scenarios. The simulation data used the negative binomial degree distribution for the upper two rows and the Poisson degree distribution for the lower two rows. The first and fourth rows are correctly matched between data and model; the first for negative binomial and the fourth for Poisson respectively. the second and third rows are mismatched between the data and the model. The first column panels denoted the result when we estimated all parameters. The second column panels denoted the result when we fixed $\beta$, and the third column is the result with $\mu$ fixed.}
	\label{fig:ga}
\end{figure}
However, the estimation failed in scenario $l_1$ and scenario $l_3$ for the mismatched Poi model, which uses only limited information. The means of the 100 posterior median are 1.76 and 1.30, respectively. Let us look at the results of fitting the model with $\beta$ or $\mu$ fixed. First, in panels (b) and (c), we can see that the accuracy of scenario $l_1+l_2$ has improved. The MSEs have been reduced by less than half. For the other three scenarios, we can see that the improvement in estimation accuracy is almost similar because the results in panel (a) were sufficiently accurate. However, if we look at the results in panels (e) and (f), which were the results with the mismatched Poi model, we can see that the accuracy is worse despite fixing $\beta$ or $\mu$ differently from the results of $R_0$. In particular, the increase in bias is noticeable. The model estimation procedure tries to fit the observed data correctly as much as possible. So, when we fix one parameter, the others compromise to fit the observed data better. So $R_0$ is relatively accurate, but $\gamma$ shows a significant bias. We can see similar bias in the estimation of $\beta$ and $\mu$ in the fitting with Poi model. 

Let's turn into estimating the summary of $\gamma$ using the data with Poisson degree distribution. When both data and model are matched (the last row in Supplementary Fig. \ref{fig:ga}), the fitting result shows accurate and precise estimation for all scenarios, regardless of fixing $\beta$ or $\mu$ or without restriction. When the model is mismatched by the NB model, the results are summarized in the third row. When all parameters are estimated, scenario $l_1 + l_2$ and scenario $l_1 + l_3$ have accurate estimations similar to that of the Poi model, but scenarios $l_1$ and $l_3$ have a little bias (panel (g)). When $\beta$ or $\mu$ is fixed, There was no significant improvement compared to no constraints.  (panel (h) and (i)). When we compare the estimation performance of the mismatched NB model to the matched NB model, MSEs are almost similar between the two cases. We have the same conclusion in the estimation of $R_0$. NB model performs better than Poi model in the estimation of $\gamma$.

\begin{table}[h!]
    \centering
    \resizebox{0.8\textwidth}{!}{  
    \begin{tabular}{c|c|c|r|r|r|r|r|r|r|r|r|r|r|r}
        \hline
        \multirow{2}{*}{\begin{tabular}{c}Degree \\ distribution \end{tabular}} & \multirow{2}{*}{\begin{tabular}{c}Fitting \\ model \end{tabular}}  & \multirow{2}{*}{Statistics} & \multicolumn{4}{c|}{No restriction} & \multicolumn{4}{c|}{$\beta$ fixed} & \multicolumn{4}{c}{$\mu$ fixed} \\ \cline{4-15}
        & & & \multicolumn{1}{c|}{$l_1$} & \multicolumn{1}{c|}{$l_1+l_2$} &  \multicolumn{1}{c|}{$l_1+l_3$} & \multicolumn{1}{c|}{$l_3$} & \multicolumn{1}{c|}{$l_1$} & \multicolumn{1}{c|}{$l_1+l_2$} & \multicolumn{1}{c|}{$l_1+l_3$} & \multicolumn{1}{c|}{$l_3$} & \multicolumn{1}{c|}{$l_1$} & \multicolumn{1}{c|}{$l_1+l_2$} & \multicolumn{1}{c|}{$l_1+l_3$} &  \multicolumn{1}{c}{$l_3$} \\ \hline
        \multirow{8}{*}{\begin{tabular}{c}Negative \\ binomial \\ degree \\ distribution\end{tabular}} 
        & \multirow{4}{*}{\begin{tabular}{c}NB \\ model \end{tabular}}
        & Mean & 0.7082 & 0.9844 & 0.9333 & 0.9084 & 0.8171 & 0.9862 & 0.9641 & 0.9506 & 0.8526 & 0.9858 & 0.9649 & 0.9446 \\ \cline{3-15}
        &  & MSE & 0.1428 & 0.0016 & 0.0095 & 0.0364 & 0.0811 & 0.0018 & 0.0070 & 0.0314 & 0.0626 & 0.0018 & 0.0079 & 0.0226 \\ \cline{3-15}
        &  & Bias$^2$ & 0.0852 & 0.0002 & 0.0045 & 0.0084 & 0.0335 & 0.0002 & 0.0013 & 0.0024 & 0.0217 & 0.0002 & 0.0012 & 0.0031 \\ \cline{3-15}
        &  & Var & 0.0583 & 0.0014 & 0.0051 & 0.0282 & 0.0481 & 0.0016 & 0.0058 & 0.0293 & 0.0413 & 0.0016 & 0.0067 & 0.0198 \\ \cline{2-15}
       & \multirow{4}{*}{\begin{tabular}{c}Poi \\ model \end{tabular}}
        & Mean & 1.7616 & 1.0018 & 0.9283 & 1.2971 & 2.0641 & 1.3066 & 1.8473 & 1.7491 & 1.9005 & 1.3269 & 1.7350 & 1.6530 \\ \cline{3-15}
        &  & MSE & 0.7128 & 0.0018 & 0.0102 & 0.2568 & 1.1461 & 0.0968 & 0.7273 & 0.5712 & 0.8209 & 0.1094 & 0.5473 & 0.4336 \\ \cline{3-15}
        &  & Bias$^2$ & 0.5801 & 0.0000 & 0.0051 & 0.0883 & 1.1322 & 0.0940 & 0.7179 & 0.5612 & 0.8109 & 0.1069 & 0.5403 & 0.4264 \\ \cline{3-15}
        &  & Var & 0.1340 & 0.0018 & 0.0051 & 0.1702 & 0.0140 & 0.0028 & 0.0095 & 0.0102 & 0.0100 & 0.0026 & 0.0071 & 0.0073 \\ \hline
        
        \multirow{8}{*}{\begin{tabular}{c}Poisson \\ degree \\ distribution\end{tabular}} 
        & \multirow{4}{*}{\begin{tabular}{c}NB \\ model \end{tabular}}
        & Mean & 0.7114 & 0.9862 & 0.9264 & 0.8383 & 0.7648 & 0.9712 & 0.9106 & 0.8749 & 0.7715 & 0.9485 & 0.8764 & 0.8356 \\ \cline{3-15}
        &  & MSE & 0.1254 & 0.0014 & 0.0100 & 0.0396 & 0.0718 & 0.0030 & 0.0124 & 0.0758 & 0.0659 & 0.0063 & 0.0206 & 0.0358 \\ \cline{3-15}
        &  & Bias$^2$ & 0.0833 & 0.0002 & 0.0054 & 0.0261 & 0.0553 & 0.0008 & 0.0080 & 0.0423 & 0.0522 & 0.0027 & 0.0153 & 0.0270 \\ \cline{3-15}
        &  & Var & 0.0425 & 0.0013 & 0.0046 & 0.0136 & 0.0617 & 0.0022 & 0.0044 & 0.0269 & 0.0138 & 0.0037 & 0.0054 & 0.0089 \\ \cline{2-15}
        & \multirow{4}{*}{\begin{tabular}{c}Poi \\ model \end{tabular}}
        & Mean & 0.9465 & 0.9946 & 0.9831 & 0.9479 & 0.9953 & 0.9916 & 0.9843 & 0.9896 & 0.9961 & 0.9911 & 0.9928 & 0.9831 \\ \cline{3-15}
        &  & MSE & 0.0236 & 0.0013 & 0.0051 & 0.0197 & 0.0051 & 0.0016 & 0.0086 & 0.0142 & 0.0036 & 0.0019 & 0.0033 & 0.0037 \\ \cline{3-15}
        &  & Bias$^2$ & 0.0029 & 0.0000 & 0.0003 & 0.0027 & 0.0000 & 0.0001 & 0.0002 & 0.0001 & 0.0000 & 0.0001 & 0.0001 & 0.0003 \\ \cline{3-15}
        &  & Var & 0.0210 & 0.0012 & 0.0048 & 0.0171 & 0.0051 & 0.0016 & 0.0085 & 0.0135 & 0.0036 & 0.0019 & 0.0033 & 0.0034 \\ \hline
    \end{tabular}}
    \caption{\textbf{Summary statistics of $\gamma$} This table summarizes the mean, MSE (mean squared error), squared bias, and variance of the 100 posterior medians of simulation according to the two fitting degree models and four scenarios. }
    \label{tb:ga}
\end{table}

\subsection*{Estimation of $\beta$}
\label{Append:beta}
\begin{figure}[t!]
	\centering
	\includegraphics[width=0.5\linewidth]{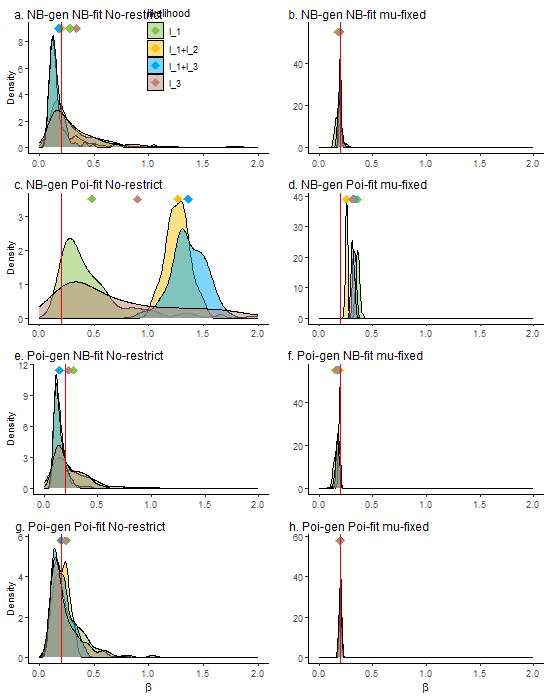}
	\caption{\textbf{Fitting summary of $\beta$.}  Each panel depicts the density of 100 posterior medians under the four scenarios. The simulation data used the negative binomial degree distribution for the upper two rows and the Poisson degree distribution for the lower two rows. The first and fourth rows are correctly matched between data and model; the first for negative binomial and the fourth for Poisson respectively. the second and third rows are mismatched between the data and the model. The first column panels denoted the result when we estimated all parameters. The second column panels denoted the result when we fixed $\mu$.}
	\label{fig:be}
\end{figure}

Supplementary Fig. \ref{fig:be} and Supplementary Table \ref{tb:be} depict the summary of the $\beta$ estimation. A different one from Supplementary Fig. for $R_0$ and $\gamma$ is the second column, which is the fitting results under fixed $\mu$. In the estimation of $\beta$ in panel (a), scenario $l_1+l_2$ and scenario $l_1+l_3$ show accurate estimates. Scenarios $l_1$ and $l_3$ show some bias, but the overall distribution contains the true value of 0.2 with large variation. Panel (c) shows the results of fitting the mismatched Poi model from the negative degree data. All scenarios failed to estimate $\beta$. Scenarios $l_1$ and $l_3$, although the data is well informed, show significant biases, with MSE values ranging from 10 to over 100 times larger than corresponding to panel (a), which results from the correctly matched NB model.  When we fixed $\mu$, matched NB model fitting results in accurate estimates in all scenarios (panel (b).). The mismatched Poi model fitting significantly improves the accuracy of the estimates (panel (d).). It still generates some bias but does not entirely fail to estimate, as shown in panel (c). The biases in the negative binomial degree are significantly mitigated by fitting the Poisson degree distribution data. The Poi model gives very accurate estimates for all scenarios (panel (g)). The NB model fitting shows a bias, but the difference is much smaller than when fitting the Poi model to negative binomial degree data. When $\mu$ is fixed, the Poi model estimates accurately and precisely in all scenarios (panel (h)), and the NB model also estimates correctly in scenario $l_1+l_2$ and scenario $l_1+l_3$. However, it shows only a little bias in scenarios  $l_1$ and $l_3$. (panel (f))

\begin{table}[t!]
    \centering
    \resizebox{0.65\textwidth}{!}{  
    \begin{tabular}{c|c|c|r|r|r|r|r|r|r|r}
        \hline
        \multirow{2}{*}{\begin{tabular}{c}Degree \\ distribution \end{tabular}} & \multirow{2}{*}{\begin{tabular}{c}Fitting \\ model \end{tabular}}  & \multirow{2}{*}{Statistics} & \multicolumn{4}{c|}{No restriction} &\multicolumn{4}{c}{$\mu$ fixed} \\ \cline{4-11}
        & & & \multicolumn{1}{c|}{$l_1$} & \multicolumn{1}{c|}{$l_1+l_2$} &  \multicolumn{1}{c|}{$l_1+l_3$} & \multicolumn{1}{c|}{$l_3$} & \multicolumn{1}{c|}{$l_1$} & \multicolumn{1}{c|}{$l_1+l_2$} & \multicolumn{1}{c|}{$l_1+l_3$} & \multicolumn{1}{c}{$l_3$} \\ \hline
        \multirow{8}{*}{\begin{tabular}{c}Negative \\ binomial \\ degree \\ distribution\end{tabular}} 
        & \multirow{4}{*}{\begin{tabular}{c}NB \\ model \end{tabular}}
        & Mean & 0.2829 & 0.1990 & 0.1797 & 0.3416 & 0.1799 & 0.1980 & 0.1954 & 0.1931  \\ \cline{3-11}
        &  & MSE & 0.0397 & 0.0247 & 0.0191 & 0.1198 & 0.0013 & 0.0001 & 0.0002 & 0.0005 \\ \cline{3-11}
        &  & Bias$^2$ & 0.0069 & 0.0000 & 0.0004 & 0.0200 & 0.0004 & 0.0000 & 0.0000 & 0.0000 \\ \cline{3-11}
        &  & Var & 0.0332 & 0.0249 & 0.0189 & 0.1008 & 0.0009 & 0.0000 & 0.0002 & 0.0004 \\ \cline{2-11}
       & \multirow{4}{*}{\begin{tabular}{c}Poi \\ model \end{tabular}}
        & Mean & 0.4829 & 1.2665 & 1.3618 & 0.8970 & 0.3611 & 0.2569 & 0.3306 & 0.3139 \\ \cline{3-11}
        &  & MSE & 0.2043 & 1.1503 & 1.3740 & 1.0071 & 0.0262 & 0.0333 & 0.0173 & 0.0132 \\ \cline{3-11}
        &  & Bias$^2$ & 0.0800 & 1.1373 & 1.3498 & 0.4859 & 0.0259 & 0.0322 & 0.0171 & 0.0130 \\ \cline{3-11}
        &  & Var & 0.1255 & 0.0131 & 0.0244 & 0.5265 & 0.0003 & 0.0001 & 0.0002 & 0.0002 \\ \hline
        
        \multirow{8}{*}{\begin{tabular}{c}Poisson \\ degree \\ distribution\end{tabular}} 
        & \multirow{4}{*}{\begin{tabular}{c}NB \\ model \end{tabular}}
        & Mean & 0.2745 & 0.1409 & 0.1479 & 0.2287 & 0.1576 & 0.1893 & 0.1769 & 0.1695 \\ \cline{3-11}
        &  & MSE & 0.0463 & 0.0056 & 0.0067 & 0.0179 & 0.0023 & 0.0005 & 0.0007 & 0.0012 \\ \cline{3-11}
        &  & Bias$^2$ & 0.0056 & 0.0035 & 0.0027 & 0.0008 & 0.0018 & 0.0001 & 0.0005 & 0.0009 \\ \cline{3-11}
        &  & Var & 0.0412 & 0.0021 & 0.0040 & 0.0173 & 0.0005 & 0.0004 & 0.0002 & 0.0003 \\ \cline{2-11}
        & \multirow{4}{*}{\begin{tabular}{c}Poi \\ model \end{tabular}}
        & Mean & 0.2492 & 0.1975 & 0.2082 & 0.2427 & 0.1998 & 0.1973 & 0.1993 & 0.1972 \\ \cline{3-11}
        &  & MSE & 0.0192 & 0.0045 & 0.0063 & 0.0275 & 0.0001 & 0.0005 & 0.0001 & 0.0001 \\ \cline{3-11}
        &  & Bias$^2$ & 0.0024 & 0.0000 & 0.0001 & 0.0018 & 0.0000 & 0.0000 & 0.0000 & 0.0000 \\ \cline{3-11}
        &  & Var & 0.0169 & 0.0045 & 0.0063 & 0.0260 & 0.0001 & 0.0005 & 0.0001 & 0.0001 \\ \hline
    \end{tabular}}
    \caption{\textbf{Summary statistics of $\beta$} This table summarizes the mean, MSE (mean squared error), squared bias, and variance of the 100 posterior medians of simulation according to the two fitting degree models and four scenarios. }
    \label{tb:be}
\end{table}

\pagebreak

\subsection*{Estimation of $\mu$}
\label{Append:mu}

\begin{figure}[t!]
	\centering
	\includegraphics[width=0.5\linewidth]{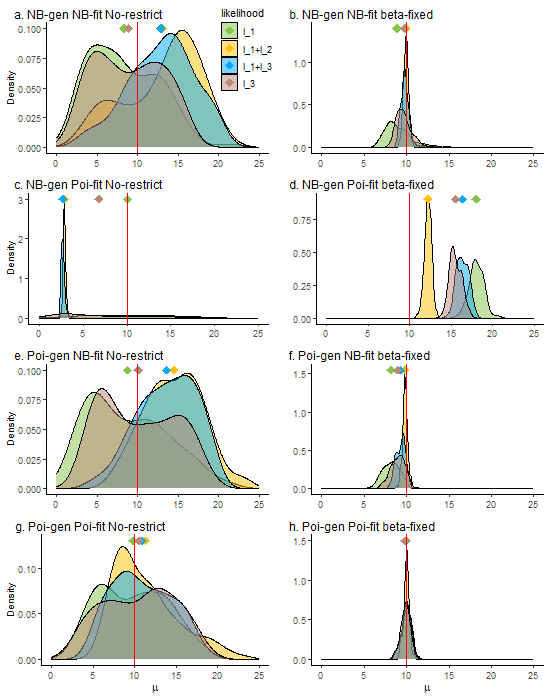}
	\caption{\textbf{Fitting summary of $\mu$.} Each panel depicts the density of 100 posterior medians under the four scenarios. The simulation data used the negative binomial degree distribution for the upper two rows and the Poisson degree distribution for the lower two rows. The first and fourth rows are correctly matched between the data and model; the first is for negative binomial, and the fourth is for Poisson, respectively. the second and third rows are mismatched between the data and the model. The first column panels denoted the result when we estimated all parameters. The second column panels denoted the result when we fixed $\beta$.} 
	\label{fig:mu}
\end{figure}

Supplementary Fig. \ref{fig:mu} and Supplementary Table \ref{tb:mu} depict the summary of the $\mu$ estimation. The second column represents the summary of $\mu$ when we fixed $\beta$. The inference of $\mu$ is relatively difficult as the observed data does not possess any direct information for the degree distribution. Looking at the results in Panel (a), we can see that the accuracy of the estimation is relatively poor across scenarios, unlike the estimation results for the other parameters. The true value of $\mu$ is 10, which is much larger in scale than the other parameters and, therefore, more variable. In practice, as the synthetic data for this simulation study does not provide direct information for a degree, the information such as infectious period or infectious time has less explanatory power in the estimation of $\mu$ than other parameters. In particular, scenario $l_1+l_2$, where we have enough information, has a worse estimation result and is even worse than scenarios $l_1$ and $l_3$. Refer to the MSEs in Supplementary Table \ref{tb:mu}. The MSE of scenario $l_1+l_2$ is 30.3378 and is the largest among the four scenarios. 

\begin{table}[b!]
    \centering
    \resizebox{0.65\textwidth}{!}{  
    \begin{tabular}{c|c|c|r|r|r|r|r|r|r|r}
        \hline
        \multirow{2}{*}{\begin{tabular}{c}Degree \\ distribution \end{tabular}} & \multirow{2}{*}{\begin{tabular}{c}Fitting \\ model \end{tabular}}  & \multirow{2}{*}{Statistics} & \multicolumn{4}{c|}{No restriction} &\multicolumn{4}{c}{$\beta$ fixed} \\ \cline{4-11}
        & & & \multicolumn{1}{c|}{$l_1$} & \multicolumn{1}{c|}{$l_1+l_2$} & \multicolumn{1}{c|}{$l_1+l_3$} & \multicolumn{1}{c|}{$l_3$} & \multicolumn{1}{c|}{$l_1$} & \multicolumn{1}{c|}{$l_1+l_2$} & \multicolumn{1}{c|}{$l_1+l_3$} & \multicolumn{1}{c}{$l_3$} \\ \hline
        \multirow{8}{*}{\begin{tabular}{c}Negative \\ binomial \\ degree \\ distribution\end{tabular}} 
        & \multirow{4}{*}{\begin{tabular}{c}NB \\ model \end{tabular}}
        & Mean & 8.3358 & 13.0521 & 12.9053 & 8.9019 & 8.8949 & 9.9252 & 9.7949 & 9.7549  \\ \cline{3-11}
        &  & MSE & 20.1339 & 30.3378 & 24.3538 & 17.7925 & 3.0956 & 0.0992 & 0.2876 & 1.3013 \\ \cline{3-11}
        &  & Bias$^2$ & 2.7695 & 9.3152 & 8.4408 & 1.2058 & 1.2213 & 0.0056 & 0.0421 & 0.0601 \\ \cline{3-11}
        &  & Var& 17.5398 & 21.2350 & 16.0736 & 16.7543 & 1.8933 & 0.0945 & 0.2480 & 1.2538 \\ \cline{2-11}
       & \multirow{4}{*}{\begin{tabular}{c}Poi \\ model \end{tabular}}
        & Mean & 10.0805 & 2.8707 & 2.7073 & 6.8221 & 18.0655 & 12.2196 & 16.3846 & 15.5517 \\ \cline{3-11}
        &  & MSE & 20.9376 & 50.8484 & 53.2332 & 30.9352 & 65.8253 & 5.1063 & 41.2833 & 31.3372 \\ \cline{3-11}
        &  & Bias$^2$ & 0.0065 & 50.8272 & 53.1838 & 10.0992 & 65.0522 & 4.9267 & 40.7634 & 30.8212 \\ \cline{3-11}
        &  & Var & 21.1426 & 0.0214 & 0.0499 & 21.0465 & 0.7809 & 0.1814 & 0.5207 & 0.5153 \\ \hline
        
        \multirow{8}{*}{\begin{tabular}{c}Poisson \\ degree \\ distribution\end{tabular}} 
        & \multirow{4}{*}{\begin{tabular}{c}NB \\ model \end{tabular}}
        & Mean & 8.8281 & 14.4943 & 13.5682 & 10.1318 & 8.1594 & 9.7167 & 9.3037 & 8.8411 \\ \cline{3-11}
        &  & MSE & 24.5207 & 32.9481 & 25.4723 & 20.7584 & 4.4233 & 1.1243 & 0.8009 & 2.7545 \\ \cline{3-11}
        &  & Bias$^2$ & 1.3734 & 20.1988 & 12.7323 & 0.0174 & 3.3879 & 0.0802 & 0.4849 & 1.3429 \\ \cline{3-11}
        &  & Var & 23.3811 & 12.8717 & 12.8716 & 20.9505 & 1.0459 & 1.0546 & 0.3192 & 1.4258 \\ \cline{2-11}
        & \multirow{4}{*}{\begin{tabular}{c}Poi \\ model \end{tabular}}
        & Mean & 9.9058 & 11.3047 & 10.8260 & 10.5699 & 9.9694 & 9.8639 & 9.8543 & 9.7554 \\ \cline{3-11}
        &  & MSE & 16.4880 & 16.3765 & 12.8447 & 18.8713 & 0.3555 & 1.1125 & 1.2951 & 1.3139 \\ \cline{3-11}
        &  & Bias$^2$ & 0.0089 & 1.7023 & 0.6824 & 0.0348 & 0.0001 & 0.0185 & 0.0212  & 0.0598 \\ \cline{3-11}
        &  & Var & 16.6456 & 14.8224 & 12.2852 & 18.7338 & 0.3590 & 1.1050 & 1.2867 & 1.2668 \\ \hline
    \end{tabular}}
    \caption{\textbf{Summary statistics of $\mu$} This table summarizes the mean, MSE (mean squared error), squared bias, and variance of the 100 posterior medians of simulation according to the two fitting degree models and four scenarios. }
    \label{tb:mu}
\end{table}

In panel (c) of the mismatched Poi model fit, Scenario $l_1+l_2$ and Scenario $l_1+l_3$ show a significant bias, while Scenarios $l_1$ and $l_3$ show a large dispersion. This imperfection in the estimation is significantly mitigated for the matched case when $\beta$ is fixed (panel (b)). The bias and variance are improved for all four scenarios resulting in accurate estimates, but the mismatched Poi model shows significant biases except scenario $l_1+l_2$ (panel (d)). This pattern is similar to Supplementary Fig. \ref{fig:be}. By overestimating $\mu$ and $\beta$ simultaneously, we end up giving relatively accurate results in the estimation of $R_0$. When we use the Poisson degree data, the matched Poi model estimated $\mu$ accurately and precisely for all scenarios regardless of fix $\beta$ (panel (h)) or not (panel (g)). When we use the mismatched NB model (panel (e)), Scenario $l_1+l_2$ and Scenario $l_1+l_3$ increase the bias, while scenarios $l_1$ and $l_3$ increase the variance. However, if we keep $\beta$ fixed (panel (f)), the NB model can estimate $\mu$ correctly. 

\end{document}